\documentstyle[aaspp]{article}
\begin{document}

\def\MSUN{\rm M_{\odot}}
\def\RSUN{\rm R_{\odot}} 
\def\MSUNYR{\rm M_{\odot}\,yr^{-1}}
\def\MDOT{\dot{M}}

\newbox\grsign \setbox\grsign=\hbox{$>$} \newdimen\grdimen \grdimen=\ht\grsign
\newbox\simlessbox \newbox\simgreatbox
\setbox\simgreatbox=\hbox{\raise.5ex\hbox{$>$}\llap
     {\lower.5ex\hbox{$\sim$}}}\ht1=\grdimen\dp1=0pt
\setbox\simlessbox=\hbox{\raise.5ex\hbox{$<$}\llap
     {\lower.5ex\hbox{$\sim$}}}\ht2=\grdimen\dp2=0pt
\def\simgreat{\mathrel{\copy\simgreatbox}}
\def\simless{\mathrel{\copy\simlessbox}}

\newcommand{\Appendix}[1]{
\appendix
\section*{Appendix #1}
\setcounter{equation}{0}
\renewcommand{\theequation}{{\rm #1}\arabic{equation}}}

\setlength{\parskip}{0.5cm}

\title{ On the role of the UV and X-ray radiation in driving 
a disk wind in X-ray binaries.}

\vspace{1.cm}
\author{ Daniel Proga, Timothy R. Kallman}
\vspace{.5cm}
\affil{LHEA, GSFC, NASA, Code 662, Greenbelt, MD 20771; proga@sobolev.gsfc.nasa.gov, tim@xstar.gsfc.nasa.gov}

\begin{abstract}

X-ray heating of the photosphere of an accretion disk is a possible mechanism 
to produce strong, broad UV emission lines in low mass X-ray binaries 
(LMXBs). However, detailed photoionization calculations show 
that this mechanism fails to produce sufficient emission 
measure at suitable ionization and optical depth. We present 
the results of hydrodynamical calculations of the disk 
photosphere irradiated by strong X-rays. We attempt to 
determine whether LMXBs can harbor significant UV-driven 
disk winds despite the effects of X-ray ionization. Such  
winds would be a likely candidate for the site of emission 
of UV lines and may better explain the observations than 
the X-ray heated disk photosphere.

We find that the local disk radiation cannot launch a wind
from the disk because of strong ionizing radiation from 
the central object. Unphysically high X-ray opacities would
be required to shield the UV emitting disk and allow
the line force to drive a disk wind.
However the same X-ray radiation that inhibits  line driving
heats the disk and can produce a hot bipolar wind or corona above the disk.
Our calculations are generally consistent with past work
on the dynamics of coronae and winds from accretion disks in LMXBs.
Additionally, our results are consistent with the UV observations
of LMXB which show no obvious spectral features associated with
strong and fast disk winds. 
To  assess the impact of X-ray heating
upon driving of a disk wind by the line force in any system
with an accretion disk we derive  analytic formulae.
In particular, we  compare results
of line-driven disk wind models for accretion disks in LMXBs and 
active galactic nuclei. The latter show  spectral features associated with
a strong and fast disk wind, the wind that has been successfully modeled
by Proga, Stone and Kallman (2000). The key parameter determining the role 
of the line force is not merely the presence of the luminous UV zone 
in the disk and the presence of the X-rays, but also the distance of 
this UV zone from the center. 
The closer the UV zone to the center, the stronger
the line force and subsequently the denser the disk wind launched
by the line force. The density of the disk wind critically determines
whether the wind will stay in a lower ionization state in the presence
of the X-ray radiation and be further accelerated by the line force
to supersonic velocities.

\end{abstract}


\section{Introduction}

Accretion disks are accepted as important components in systems
ranging from cataclysmic variables (CVs) and protostars to 
active galactic nuclei (AGN).
These accretion disks can produce intense radiation fields over
a  wide range of wavelengths, with the most energetic
photons and most of the flux emitted in the innermost part of the disk. 
The photon energy and the flux decrease with increasing radius.  
Detection of this radiation is the most secure observational evidence for 
the existence of disks.  

X-ray binaries, in particular low mass X-ray binaries
(LMXBs), are systems which crudely resemble CVs 
in their system parameters: masses, accretion rates and orbital separation.
However, in LMXBs the detection of disks is complicated by the fact
that the radiation from the innermost regions, near the compact object,
is so intense that it dominates the bolometric luminosity, and 
the customary intrinsic disk radiation is overwhelmed by 
reprocessed X-rays (Shakura \& Sunyaev 1973; Pacharintanakul \& Katz 1980;
Meyer \& Meyer-Hoffmeister 1982).  As a result, direct evidence 
for the existence of disks in these systems is less secure than in 
CVs.  Studies of the ultraviolet (UV) spectra and 
the UV and optical to X-ray intensity ratios of several LMXBs indicate 
that beyond $\sim10^9$ cm from disk center, X-ray reprocessing should 
dominate over power generated locally by viscous dissipation 
(e.g., van Paradjis 1983; Blair et al. 1984; Vrtilek et al. 1990,
1991; Kallman, Raymond \& Vrtilek 1991).
Models for heated disks are consistent 
with the few UV fluxes and spectra which have been observed 
(Vrtilek et al., 1990; Kallman, Boroson and  Vrtilek 1998);
the observed UV fluxes from LMXBs exceed those predicted by 
viscous dissipation alone by a factor $\simless$10.

One aspect of LMXB spectra which remains a puzzle is the origin of the UV emission
lines, such as C~IV $\lambda$1550 and N V $\lambda$1240.  These lines
typically have luminosities of $\sim 10^{31}$ erg s$^{-1}$ or greater, 
and appear single peaked but with widths $\sim$100 km s$^{-1}$ 
(e.g. Kallman, Boroson and  Vrtilek 1998; Vrtilek et al., 1994).  
It is straightforward to  show that if these lines are emitted via 
collisional excitation from 
optically thin gas, then their fluxes alone imply an emission measure
of n$^2$V$\sim 10^{58}$ cm$^{-3}$ or greater, assuming cosmic abundances and 
favorable ionization balance.  Past work has tested the hypothesis that this 
originates from an X-ray heated disk chromosphere, but this 
fails to provide sufficient emission measure at suitable ionization 
and optical depth (Raymond, 1993; Ko and Kallman, 1994).  The most
obvious reason for this failure is that only a very thin layer on the 
heated disk is optically thin to incident X-rays and to reprocessed 
UV lines.  

Theoretical models predict that X-ray heating can have profound effects on 
the disk dynamics and vertical structure.  Since X-rays tend to heat low 
density gas to a temperature $T_{IC}\sim 10^7$K, the surface of a heated 
disk is expected to either puff up and form a static corona, or to emit 
a thermal wind, depending whether the thermal velocity exceeds the local 
escape velocity, $v_{esc}$ 
(e.g. Begelman, McKee and Shields, 1982, hereafter BMS).  
Since the 
local escape velocity decreases with radius, the inner disk corona 
is expected to remain bound and static.  
The photons scattered by the corona can
explain the gradual and partial eclipse of X-rays at optical minimum
when the companion star blocks the line of sight of the compact object
in certain 
nearly edge-on systems (White \& Holt 1982; McClintock et al. 1982).
The boundary between the static corona and wind region occurs at a radius
which is comparable to the disk outer radius, so that the effects of 
the thermally-driven wind on the disk mass budget and X-ray spectrum 
and light curve are expected to be small.  A hot, X-ray heated wind is likely
to have negligible effect on the  appearance of an LMXB 
in the UV continuum or lines.

A ubiquitous phenomenon in cataclysmic variable UV spectra is the 
appearance of P Cygni profiles in the strong resonance lines such as 
C~IV $\lambda$1550 and N V $\lambda$1240.  A possible explanation for this 
is a wind from the accretion disk driven by UV radiation pressure, 
analogous to the winds in OB stars (e.g. Cordova and Mason, 1982;
Drew 1987).  Theoretical models have demonstrated that this mechanism 
can account for many of the features of the observed lines, such as 
inclination angle dependence, and velocity width (
Proga, Stone and Drew, 1998,
hereafter PSD~98; Pereyra Kallman and Blondin, 2000). However
the radiation-driven disk wind model cannot  account for all
of the features, in particular, relatively high mass loss rates observed 
in CVs. Comparison of 
mass loss rates predicted by the  models with observational 
constraints show that either mass accretion rates in high-state
CVs are higher than presently thought  by a factor of 2-3  or that radiation
pressure alone is not quite sufficient to drive the observed hypersonic 
flows (Drew \& Proga 2000; Mauche \& Raymond 2000).

In LMXBs, UV driven winds are not expected to be efficient, 
owing to the effects of X-ray ionization; wind material exposed to a 
strong flux of X-rays will become highly ionized and will therefore 
have a lower concentration of ions capable of providing UV line opacity
and radiative driving.  However, it is possible that the column density 
of the wind material is sufficient to allow a region which is 
shielded from X-ray ionization and which can therefore support a 
UV-driven wind.  Such a wind can potentially account for the observed 
properties of the UV lines:  if the wind velocity field contains a 
significant poloidal component then the line profiles will not be 
double-peaked, and the wind density may be low enough and the 
velocity gradient sufficiently large to avoid line self-shielding.

UV driven winds are a likely explanation for the outflows in 
AGNs, particularly broad absorption line (BAL) quasars.  These 
objects resemble LMXBs in their X-ray/bolometric ratios, 
and recent simulations have demonstrated that AGN disk winds 
can provide sufficient shielding against X-ray ionization to allow 
significant mass loss near the disk plane (Proga, Stone and Kallman 2000, 
hereafter PSK).
In this paper we attempt to determine whether stellar mass systems with 
strong X-rays, i.e. LMXBs, can harbor significant UV-driven 
disk winds in a manner analogous with AGN disk winds.  If so, 
we would consider such winds to be a likely candidate for the 
site of emission of UV lines.  In doing so, we use many of the same 
assumptions and numerical techniques applied to our study of 
AGN disk winds in PSK.

The outline of this paper is as follows: in Section~2 we discuss previous work 
on LMXB disk winds. In Section 3, we describe our 
computational algorithm. We present our results in Section~4 and discuss them together
with perceived limitations  in Section~5. In Section~5 we also
discuss the limits on the validity of the analogy between LMXBs and AGNs.  
We summarize our conclusions in Section~6.

\section{Past Work}

Past work on the  dynamics of coronae and winds from 
accretion disks in LMXBs has concentrated exclusively on 
the effects of X-ray heating, beginning with the basic work by BMS and 
Begelman \& McKee (1983).  With a few exceptions, treatment of  
this inherently two-dimensional problem was reduced to one dimension by
solving the vertical structure on cylinders at different radii.
Such a simplified treatment ignores the effects of radiative transfer in 
the corona. To address this problem, 
Ostriker, McKee \& Klein (1991) calculated the flux incident at the
base of the corona for an isothermal X-ray heated corona in 
the single-scattering limit while Murray et al. (1994) used a two-dimensional
flux limited radiative diffusion technique. To predict the iron
K-shell feature for X-ray binaries, Vrtilek, Soker \& Raymond (1993) performed
two-dimensional Monte-Carlo simulations for the radiative transfer.

The first two-dimensional hydrodynamical calculations of the structure of 
a thermally-driven wind from
an accretion disk heated by X-rays from a central source were performed
by Woods et al. (1996). Earlier two-dimensional calculations of the X-ray 
heated wind did not include any of the
disk in the computational domain (Melia, Zylstra \& Fryxell 1991; Melia \& 
Zylstra 1992; Balsara \& Krolik 1993). 
Woods et al.'s (1996) calculations were designed to understand the types of 
wind solutions available for different luminosities and Compton temperatures.
Their calculations help to extend and improve on the analytic
predictions of BMS for X-ray heated winds by including non-Compton processes
such as photoionization heating and line cooling. The results from 
these improved two-dimensional calculations
agree well with many  predictions given by BMS, even when non-Compton 
processes dominate.

Other mechanisms that can drive a wind
off the disk photosphere have been studied primarily in the context 
of CVs and AGNs.  Examples include magneto-centrifugal force due to 
an accretion disk (e.g., Blandford \& Payne 1982) and radiation pressure due 
to lines  (e.g., Vitello \& Shlosman 1988; Murray et al. 1995). 
The relative importance of these different mechanisms is likely to depend 
on the type of the system. For example, in systems with cool disks, such 
as low mass young stellar 
objects (YSO),  the magneto-centrifugal force is the only promising
mechanism because radiation pressure due to lines (line force, in short) and 
thermal expansion are likely to be negligible (e.g., K\"onigl \& Ruden 1993). 
On the other hand, the line force is a candidate for wind driving in systems 
with hotter disks such as  in non magnetic CVs (e.g., PSD~98, Proga~2000).  
For example, in the most favorable conditions (i.e., high
UV flux and low X-ray flux) the line force can exceed the radiation
pressure due to electron scattering by a factor as high as $\sim 2000$
(e.g., Castor Abbott, \& Klein 1995, hereafter CAK; Abbott 1982; Gayley 1995).
Thus systems with luminosity as low as 0.1 per cent of their
Eddington limit can produce a powerful high velocity wind.
Magneto-centrifugal winds cannot be excluded from such systems, but they may  
not be required  to account for the observed lines.
When wind material is exposed to the  UV  radiation but also  
to strong X-rays, the conditions for line driving can dramatically 
worsen as X-rays may highly ionize the gas and reduce a concentration of ions
capable of providing UV line opacity.

Models for wind driving in X-ray binaries with massive companions (MXBs)
have shown that it is possible that  line-driven winds
can survive against X-ray ionization.  For example,
Stevens \& Kallman (1990) and (Stevens 1991) calculated the structure of 
line-driven winds, including the effects of the ionization balance on the 
UV line opacity, for various X-ray luminosities with and without optical depth effects.
Additionally, they considered irradiation at various directions including
the normal to the mass losing photosphere. In none of the models did
the irradiation penetrate below the so-called wind critical point
where the wind is already launched and its mass loss rate is determined.  
These models verified that the irradiation can 
severely decrease the wind velocity in the supersonic portion of the flow but 
does not greatly affect the wind mass loss rate. The line-driven  wind from a disk 
has an advantage over the wind from a star in surviving against X-ray ionization
owing to the fact that the X-ray flux strikes the disk always at a grazing angle whereas
it can strike the star nearly normal to the photosphere.
Thus it is possible that the X-rays do not irradiate the disk but rather 
the line-driven disk wind. Such a case has recently been found by us (PSK), 
in which we calculated axisymmetric time-dependent 
hydrodynamic calculations of line-driven winds from accretion disks
in AGN.   The central engine of  the AGN is a source of both ionizing X-rays and
wind-driving  UV photons. To calculate the radiation force, we took into account radiation
from the disk and the central engine. The gas temperature and 
ionization state in the wind were calculated 
self-consistently from the photoionization and heating rate of the central
engine.  We found that the X-rays 
from the central object are significantly attenuated by the disk
atmosphere so they 
cannot prevent the local disk radiation from pushing matter
away from the disk. 
However in the supersonic portion of the flow high above the disk, 
the X-rays can overionize the gas and decrease the wind terminal velocity.

As mentioned in Section~1,
X-ray irradiation has been suggested as a mechanism to boost the UV radiation from 
the disk in LMXBs at radii $\simgreat 10^9$~cm. 
UV radiation is known to be effective
in driving a wind  because there are many  spectral
lines ($\sim 10^5$) to scatter UV continuum photons provided that the gas
is at temperature $\simless 50, 000$ (e.g., CAK; Abbot 1982; Puls et al. 2000).  
Thus we expect that if
the line force is important in driving disk winds
it should be for radii from a few $10^9$~cm to $\sim 10^{10}$~cm
because most of the disk UV radiation (generated by local disk viscosity
and due to X-ray reprocessing) is emitted at those radii.
We also note that thermal expansion in this region of the disk
cannot  produce  a robust wind but rather a nearly isothermal 
corona for parameters suitable for LMXBs (BMS, Woods et al. 1996, and
Section~4).

\section{Disk outflows driven by radiation}

Our goal in this work is to understand how inclusion of the line force
can change the dynamics of a corona/wind from the X-ray heated disk in a LMXB.
We use the model from PSK and simply adopt model parameters
suitable for a typical LMXB.   We compare our results with results
from BMS and Woods et al. for the pure  X-ray heated disks and 
with the results
for the pure line-driven disk winds from PSD~98, Proga, Stone \& Drew
(1999, hereafter PSD~99), Proga (1999) and PSK.

Our 2.5-dimensional hydrodynamical numerical method is in most respects
as described by PSK. Here we only describe the key elements of the method
and refer a reader to PSK for details.

We study the dynamics of a wind from a   Keplerian, geometrically-thin and
optically-thick disk. To calculate the structure and evolution of a wind, 
we solve the equations of hydrodynamics
\begin{equation}
   \frac{D\rho}{Dt} + \rho \nabla \cdot {\bf v} = 0,
\end{equation}
\begin{equation}
   \rho \frac{D{\bf v}}{Dt} = - \nabla P + \rho {\bf g}
 + \rho {\bf F}^{rad} 
\end{equation}
\begin{equation}
   \rho \frac{D}{Dt}\left(\frac{e}{ \rho}\right) = -P \nabla \cdot {\bf v} 
   + \rho \cal{L}  ,
\end{equation}
where $\rho$ is the mass density, $P$ is the gas pressure, 
${\bf v}$ is the velocity, $e$ is the internal energy density,
$\cal{L}$ is the net cooling rate,
${\bf g}$ is the gravitational acceleration of the central object,
${\bf F}^{rad}$ is the total radiation force per unit mass.
We assume a fully ionized gas for which $P~=~(\gamma-1)e$, and we 
consider models with $\gamma=5/3$. 
In solving the energy equation (eq. 3), we  include
absorption and emission of radiation through the radiation net cooling rate, 
$\cal L$.

We perform our calculations  in spherical polar coordinates
$(r,\theta,\phi)$. We assume, natural for a Keplerian disk, 
axial symmetry about the rotational axis
of the disk ($\theta=0^o$). 
However we allow the $\theta=90^o$ axis
to be above the disk midplane, at the height, $z_o$, corresponding
to the disk pressure scale height. 

A realistic description of the radiation field from the interface between
the disk and the neutron star
would require detailed knowledge of the geometry of the flow near the
neutron star, which is beyond the scope of this investigation. 
Instead we model the innermost part of the disk as a point source
of radiation located at the origin of our computational grid
(that is located a height $z_o$ above the disk midplane). This is
equivalent to assuming the central source has  finite width.

Our computational domain is defined to occupy the radial range
$r_i~=~10^3~r_\ast \leq r \leq \ r_o~=~ 10^4~r_\ast$, where $r_\ast$ is
the  radius of the neutron star,  and the angular range
$0^o \leq \theta \leq 90^o$.   This choice is determined by the radial 
temperature distribution of the disk; for $r~\leq ~10^3~r_\ast$ the disk 
temperature is predicted to be $> 10^5$K, and winds at temperatures 
greater than this have a greatly reduced efficiency of radiation
pressure driving (Abbott, 1982).
Our gridding of the computational domain and 
the boundary conditions are as in PSK. Here we only summarize our
boundary conditions.
At $\theta=0$, we apply an axis-of-symmetry boundary condition.  For
the outer radial boundary, we apply an outflow boundary 
condition.  For the inner radial boundary $r=r_{\ast}$
and for $\theta=90^o$, we apply reflecting boundary conditions for the density,
velocity and internal energy.
To represent steady conditions in the photosphere at the base of the wind, 
during the evolution of each model we apply the constraints that in
the first zone above the $\theta=90^o$ plane the radial velocity $v_r=0$,
the rotational velocity $v_\phi$ remains Keplerian, the density is
fixed at $\rho = \rho_0$ at all times and also the internal energy
is fixed at $e=e_o=\frac{\rho_o k T_D}{\mu m_p (\gamma -1)}$ at all times
($m_p$ is the proton mass, and $\mu$ is the mean molecular weight).
Also as in PSK, we solve eqs. 1-3 using an extended version of 
the ZEUS-2D code (Stone \& Norman 1992). 

The geometry and assumptions needed to compute the radiation
field from the central object are similar to those described in PSK. 
We exactly follow PSK to compute the radiation from the disk 
when we consider a flat disk. However we also consider 
a slightly concave surface of the disk and then we follow Vrtilek et al. 
(1990) to compute the effective temperature of the irradiated disk.
The disk photosphere is  coincident with the $\theta=90^o$ axis.
We specify the radiation field of the non-irradiated disk  by assuming 
that the temperature 
follows the radial profile of the optically thick accretion disk 
(Shakura \& Sunyaev 1973), and therefore depends on
the mass accretion rate in the disk, $\dot{M}_a$, the mass of the neutron 
star, $M_{NS}$  and  the inner edge of the disk, $r_\ast$, that coincides
with the  radius of a neutron star. We include the effect of a radiant
central object both in modifying the radial temperature of the disk,
and in providing a contribution to the driving radiation field.

We express the central object 
luminosity $L_\ast$ in units of the disk luminosity $L_\ast=x L_D$.
We define the central object luminosity as the intrinsic luminosity
of a neutron star, $L_{NS}$  plus the luminosity
due to release of accretion energy outside the disk proper, for example, in 
the boundary layer between the disk and the star, $L_{BL}$. 
If the latter luminosity dominates then x=1 because half of the total 
accretion energy is release in the disk and the other half in the boundary 
layer (e.g., Shakura \& Sunyaev 1973). We expect that this is the case
in LMXBs because these systems have relatively high mass accretion rates
(i.e., $L_{NS} < L_{LB}\approx L_D$).
As in PSK, we allow for the situation
when only some fraction of the central object luminosity 
takes part in  driving a wind via lines. We identify this fraction 
as the luminosity in the UV band, $f_{\rm UV}L_\ast$. 
We refer to the fraction of the luminosity that is responsible 
for  ionizing  the wind to a very high state as the luminosity in 
the X-ray band, $f_{\rm X}L_\ast$. For simplicity, we assume here
that this fraction of the luminosity does not contribute to line driving. 
We call the luminosity in the remaining bands, mainly optical and infrared, 
as $f_{\rm O,IR} L_\ast$. We assume that $f_{\rm O, IR}L_\ast$
is the part of the luminosity that does not change the dynamics 
or ionization of the wind. 
We consider relatively high accretion rates that imply a high temperature
($> 10^7$~K) of the central object. Therefore 
we set $f_{UV}$ and$f_{\rm OPT,IR}$ to zero in the remaining part of 
the paper. In turn, we assume that $f_{\rm X}=1$.
We take into account the irradiation of the disk by the central object, 
assuming that the disk re-emits all absorbed energy locally 
and isotropically. 
We note that for a flat disk, the contribution from irradiation is negligible 
for $x\sim 1$ and large radii. The contribution
from irradiation can be significant when the disk height increases with radius
(see below).

We approximate the radiative acceleration due to lines
using a modified CAK method.  The line force
at a point defined by the position vector $\bf r$ is
\begin{equation}
{\bf F}^{rad,l}~({\bf{r}})=~\oint_{\Omega} M(t) 
\left(\hat{n} \frac{\sigma_e I({\bf r},\hat{n}) d\Omega}{c} \right)
\end{equation}
where $I$ is the frequency-integrated continuum intensity in the direction
defined by the unit vector $\hat{n}$, and $\Omega$ is the solid angle
subtended by the disk and central object at the point. 
The term in brackets is the electron-scattering radiation force,
$\sigma_e$ is  the mass-scattering coefficient for free electrons,
and $M(t)$ is the force multiplier -- the numerical factor which
parameterizes by how much spectral lines increase the scattering
coefficient. In the Sobolev approximation, $M(t)$ is a function
of the optical depth parameter
\begin{equation}
t~=~\frac{\sigma_e \rho v_{th}}
{ \left| dv_l/dl \right|},
\end{equation}
where $v_{th}$ is the thermal velocity, 
and $\frac{dv_l}{dl}$ is the velocity gradient along the line of sight, 
$\hat{n}$.

To calculate the force multiplier, we adopt the CAK analytical expression  
modified by Owocki, Castor \& Rybicki (1988, see also PSD~98)
\begin{equation}
M(t)~=~k t^{-\alpha}~ 
\left[ \frac{(1+\tau_{max})^{(1-\alpha)}-1} {\tau_{max}^{(1-\alpha)}} \right]
\end{equation}
where $k$ is proportional to the total number of lines, $\alpha$ is the 
ratio of optically-thick to optically-thin lines, 
$\tau_{max}=t\eta_{max}$ and $\eta_{max}$ is a parameter determining
the maximum value, $M_{max}$ achieved for the force multiplier.  
Equation 11 shows the following limiting behavior:
\begin{eqnarray}
\lim_{\tau_{max} \rightarrow \infty}~M(t) & = & k t^{-\alpha} \\
\lim_{\tau_{max} \rightarrow 0}~M(t) & = & M_{max},
\end{eqnarray} 
where $M_{max} = k (1-\alpha)\eta_{max}^\alpha$. The maximum value of the 
force multiplier is a function of physical parameters of the wind and
radiation field. Several studies have showed that $M_{max}$ is roughly
a few thousand for gas ionized by a weak or moderate radiation field
(e.g., CAK; Abbott 1982; Stevens \& Kallman 1990; Gayley 1995).
As the radiation field becomes stronger and the gas becomes more ionized
the force multiplier decreases asymptotically to zero.

We evaluate the radiation force in four steps. First, we calculate the
intensity, the velocity gradient in the $\hat{n}$ direction and then the 
optical depth parameter $t$. Second, we calculate the parameters of the
force multiplier using a current value of the photoionization
parameter:
\begin{equation}
\xi = \frac{4 \pi {\cal F}_{\rm X}}{n},
\end{equation}
where ${\cal F}_{\rm X}$ is the local  X-ray flux, $n$
is the number density of the gas ($={\rho}/({m_p \mu })$). 
In this step we adopt results of Stevens \& Kallman (1990). 
To estimate $\xi$, we correct the ionizing flux for the optical 
depth effects: ${\cal F}_{\rm X}= exp(-\tau_{\rm x}) L_X /(4 \pi r^2)$, where 
$L_x$ is the luminosity of a point-like X-ray source and $\tau_{\rm x}$
is the optical depth between the source and a point in a wind
(see PSK for more details). Thus our approach differs from 
Stevens \& Kallman's one, who estimated $\xi$ assuming that the gas
is optically thin (i.e., ${\cal F}_{\rm X}= L_X /(4 \pi r^2$).
Then we calculate the radiation force exerted by radiation along $\hat{n}$. 
Third, we integrate the radiation
force over the solid angle subtended by the radiant surface. 
Finally, we correct the radiation force in the radial direction
for the optical depth effects; we use
the absorption coefficients that are
representative for the $UV$ and $X$ bands, $\kappa_{UV}$ and $\kappa_{X}$,
respectively.

Our numerical algorithms and methods to evaluate the radiation force
are described in PSK and PSD~99.  
Below we describe our calculations of the intensity of the irradiated
disk that differ from PSK's calculations when we consider  
a concave disk.

For a flat disk, in the disk plane at $r=r_{D}$,  
the total isotropic disk intensity is:
\begin{equation}
I_{D,total}(r_D) = I_{D}(r)+ I_{irr}(r),
\end{equation}
where $I_D$ is the intrinsic disk intensity and  
$I_{irr}$ is the contribution to the radiation emitted by the disk
which results from irradiation by the central object.
According to the standard steady disk model (e.g., Pringle 1981)
\begin{equation}
I_{D}(r_D) =~\frac{3 G M_{NS} \MDOT_a}{8 \pi^2 r_\ast^3} 
\frac{r_\ast^3}{r_D^3}\left(1 -\left( \frac{r_\ast}{r_D}\right)^{1/2}\right). 
\end{equation}
The contribution from the irradiation for a flat disk is
\begin{equation}
I_{irr}(r) = \frac{3 G M_{NS} \MDOT_a}{8 \pi^2 r_\ast^3}
\frac{x}{3\pi}\left(\arcsin \frac{r_\ast}{r_D} - 
\frac{r_\ast}{r_D} \left(1 - 
\left(\frac{r_\ast}{r_D}\right)^2\right)^{1/2}\right)
\end{equation}
(e.g., PSD~98).  
For a flat disk at large radii and $x \sim 1$, 
the contribution from the irradiation is negligible compared to the intrinsic 
disk intensity ($I_{irr}<< I_D$). Such a case was considered by PSK.
However, the contribution from the irradiation can be dominant
if we assume that the disk height, $h$ increases with radius (e.g.,
Cunnigham, 1976).
For example, if $h \propto r^n$, and $n=9/7$ then the disk intensity is
\begin{equation}
I_{irr}(r) = \frac{\sigma}{\pi} \left(\frac{f_2 f_3 \MDOT_a}
{14 \pi \sigma r_\ast} \left(\frac{f_1 k G M_{NS}}{\mu m_p r_D^3}
\right)\right)^{8/7},
\end{equation}
where $f_1$ is a factor of order of unity, the exact value depends
on the details of the vertical structure of the disk, $f_2$
is the absorbed fraction of the radiation impinging on the surface and
$f_3$ is a factor of unity, takes account of possible anisotropy of 
the radiation (Vrtilek et al. 1990). Other symbols have their
usual meaning. Here we use the following values for the different
parameters: $\mu=1$, $f_1=1$, $f_2=0.5$ and $f_3=0.5$. Adopting
now values of the remaining parameters typical for a LMXBs (see Section 4)
we find that the contribution from the irradiation will dominate
over the contribution from the intrinsic disk intensity 
(i.e., $I_{irr}/I_{D}\simless 3$) for
$r_d \simgreat  10^9$~cm where the disk temperature, 
$T_D\equiv (\pi I_D/\sigma)^{1/4} < 10^5$~K. Thus the optical and UV fluxes
are dominated by reprocessed flux. We limit our calculations  to radii
where the irradiation is strong enough to change the conditions in 
the upper atmosphere of the disk (e.g. the disk effective temperature) but 
too weak to change the disk internal structure (e.g., 
the midplane temperature, $T_m$ and therefore the disk thickness).
For the irradiation to change the disk structure, the radiation temperature
of the surface of a irradiated disk, $T_{irr}$ should exceed the midplane
temperature. The latter is given by radiative equilibrium as $T^4_m \sim \tau_D
T^4_D$, where $\tau_D$ is the optical depth of the disk. 
We consider the optically thick standard disk with $\tau_D> 40$ 
using our disk parameters (e.g., Frank, King, \& Raine 1985).
We expect then that the irradiation does not alter  the disk structure
because $T^4_{irr}/T^4_D=I_{irr}/I_D< \tau_D$.

We limit our presentation and discussion of the results to the models where 
the disk intensity was calculated assuming a concave disk (eqs 9, 10 and 12). 
As we mentioned above, in such a case the UV flux is much higher than for 
a flat disk. Therefore we can use the results for the concave disk
as an upper limit to the results for a flat disk. 

We calculate the gas temperature  assuming that the gas is optically thin to 
its own cooling radiation and that the abundances are cosmic (Withbroe 1971).
The net cooling rate depends on the  density, $\rho$, the temperature, $T$, 
the ionization parameter $\xi$, and the characteristic temperature of 
the X-ray radiation $T_{\rm X}$. It is then possible to fit analytical 
formulae to the heating and cooling rate obtained from  detailed  
photoionization calculations for various $T_{X}$, $T$, and $\xi$ (e.g., 
Blondin 1994). We use Blondin's result to express the net cooling  
rate $\cal {L}$ in equation 3 (see PSK). 

\section{Results}

We specify our models by the following parameters:
In all our calculations we assume the mass of the non-rotating neutron star,  
$M_{NS}=1.4~\rm \MSUN$ and its  radius, $r_\ast=10^{6}$~cm.   
To determine the radiation field from the disk, we assume the mass accretion
rate $\MDOT_a=2\times10^{-9}$~M$_{\odot}$~yr$^{-1}$ or  
$2\times10^{-8}$~M$_{\odot}$~yr$^{-1}$. 
These system parameters yield the disk Eddington number, $\Gamma_D=0.067$,
and 0.67, respectively. We define the disk Eddington number as the disk
luminosity, $L_D = (M_{NS}~\MDOT_a G)/(2 R_{NS})$ in units of the
Eddington limit for the non-rotating accreting star, 
$L_{\rm Edd}= 4 \pi c G M_{NS}/\sigma_e$ (i.e., 
$\Gamma_D\equiv L_D/L_{\rm Edd}= (\sigma_e \MDOT_a)/(8\pi c r_{NS}$)).
We consider the disk irradiation for 
the case of a concave disk ($n=9/7$) and use equation 9 
combined with eqs. 10 and 12 to calculate the disk intensity.
Thus the disk intensity in our computational
domain is dominated by the contribution from X-ray reprocessing (Section~3).
We set the offset of the computational domain, $z_o~= H_D \approx
3 \Gamma_D~r_\ast$.
The radiation field from 
the central object is specified by the additional parameter: $x=1$. 
To calculate the line force we 
adopt the force multiplier parameter $\alpha=0.6$. We calculate the 
remaining parameters of the multiplier, i.e., $k$ and $\eta_{max}$ 
as functions of  the photoionization parameter, $\xi$ using  eqs 15 and 16 in 
PSK. 
The force multiplier depends only formally on the thermal speed, 
$v_{th}$ which we set to 20 ${\rm km~s^{-1}}$, i.e., the thermal speed of 
a hydrogen atom at the temperature of 25000~K (Stevens \& Kallman 1990).  
To calculate the gas temperature, we assume the temperature of 
the X-ray radiation, $T_{\rm X}=10^8$~K and the line cooling parameter 
$\delta=1$ (see Blondin  1994 and PSK).
The attenuation of the UV radiation is calculated using
$\kappa_{\rm UV}= 0.4~{\rm g^{-1}~cm^2}$ for all $\xi$.
The resulting optical depth corresponds to the Thomson optical depth.
To study the role of the X-ray attenuation, 
we explore various prescriptions for $\kappa_{\rm X}$.
Table~1 specifies the parameter values of the models discussed below.

Our computational domain in the radial direction (i.e., 
$r_i~=~10^9~{\rm cm} \leq r \leq \ r_o~=~10^{10}~{\rm cm}$)
covers the part of the disk where the effective temperature of the 
disk, $T_D$ is between $\sim 10^5$~K and $\sim 10^4$~K. We assume then
that all the radiation emitted by this region of the disk can drive a wind via
lines, regardless of whether the radiation is generated locally by viscous 
dissipation or the radiation is due to X-ray reprocessing.
We note that for our parameters, the so-called Compton radius
($R_{IC}\equiv G M_{NS} \mu m_p/k T_{IC}\approx 9\times10^{10}$~cm, where $T_{IC}$ 
is the Compton temperature and other symbols have their usual meaning)
(e.g., BMS) is larger than the outer radius of our computational domain
and larger than the orbital separation of most LMXBs.

Figure~1 presents the results from the run~H2 with 
$\MDOT_a=2\times10^{-8}$~M$_{\odot}$~yr$^{-1}$ (see Table~1). For this model,
we computed the attenuation of the X-ray radiation assuming 
$\kappa_{\rm X}=40~{\rm g^{-1}~cm ^2}$ for $\xi \leq 10^5$ and  
$\kappa_{\rm X}=0.4~{\rm g^{-1}~cm ^2}$ for $\xi > 10^5$.
The figure shows the instantaneous density, temperature and photoionization
distributions and the poloidal velocity field of the model.
The flow settles quickly into a steady state. Our calculation contains
(i) a hot, very low density flow in the polar region
(ii) a hot, dense and fast outflow from the disk and (iii) a warm
zone near the disk midplane that is nearly in hydrostatic equilibrium.

In the polar region, the density is very small and close to the lower limit
we set on the grid, i.e., $\rho_{min}=10^{-22}~\rm g~cm^{-3}$.
The gas temperature in the polar region is close to the Compton temperature 
($\sim 0.25 T_{X}$). For such high temperature (and high photoionization
parameter) the line force is negligible in the polar region.

The wind from the disk is much denser 
than the gas in the polar region but it has a very similar temperature. 
For $r = r_i$, at the wind base, $\rho \sim 10^{-9}~\rm g~cm^{-3}$ 
and the gas temperature is $\sim 10^7$~K.
We do not resolve the transition between the X-ray heated gas and 
the unheated gas. The gas conditions are different just
one grid point, in the $\theta$ direction, below the wind base.
For example, the gas temperature assumes the disk effective temperature, 
$2\times 10^5$~K, and the density is $\rho \sim 10^{-7}~\rm g~cm^{-3}$.
The photoionization parameter in the disk wind is many orders of magnitude 
lower than in the polar region yet it is still very high ($\simgreat 10^5$)
which implies full ionization of the gas in the wind. The wind then
is not driven by the line force but by the combination of the
thermal expansion and the radiation force due to electrons.
We calculated a test model where we  switched off the X-ray heating and a test
model where we switched of the radiation force due to electrons. 
In both tests we found no disk wind
but a complex and slow disk corona that is hot when the radiation force
was switched off and warm ($T~<~10^5$~K)
when the X-ray heating was switched off.
The poloidal velocity (Figure 1, the bottom right panel) shows that the
gas has a significant radial component right from the wind base.

Figure~2 presents the run of the density, radial velocity, mass flux density,
accumulated mass loss rate, photoionization
parameter and column density as a function of  the polar angle, $\theta$, at
the outer boundary, $r=r_o=10^{10}$~cm from Figure~1.
The accumulated mass loss rate is given by:
\begin{equation}
d\dot{m}(\theta) =
4 \pi r_o^2 \int_{0^o}^\theta \rho v_r \sin \theta d\theta,
\end{equation}
while
the column density is given by:
\begin{equation}
N_H(\theta)= \int_{r_i}^{r_o} \frac{\rho(r, \theta)}{\mu m_p } dr.
\end{equation}
The gas density increases strongly with $\theta$ between $0^o$ and $45^o$.
Then the density increases gradually from $\sim 10^{-13}~\rm g~cm^{-3}$
at $\theta=45^o$ to  $\sim 10^{-11}~\rm g~cm^{-3}$ at $\theta=87^o$.
The moderate increase of the density in the wind is followed by a
dramatic increase for $\theta > 87^o$, as  might be expected
of a density profile determined by hydrostatic equilibrium.
The radial velocity has 
a very broad
peak for $10^o \simless \theta \simless 40^o$ with 
the maximum of $2100$~km~$\rm s^{-1}$ at $\theta=35^o$.

The accumulated mass loss rate is negligible for $\theta \simless 40^o$
because of the very low gas density and despite the fact that the velocity
is the highest in this range of the polar angle.
For $\theta \simgreat 40^o$, $d\dot{m}$ increases gradually 
to $\sim 1\times 10^{17}~\rm g~s^{-1}$ 
at $\theta \approx 88^o$.  Then, in the region close to 
the $\theta=90^o$ axis, where the gas density starts to rise very sharply and 
where the motion is typically slow and more complex, the accumulated mass 
loss rate is subject to fluctuations. It may even be negative for 
$\theta \simless 90^o$ where it is dominated by the contribution from 
the slow-moving region very close to the disk -- a contribution 
that is very markedly time-dependent. The steady increase of the accumulated 
mass loss rate with $\theta$, for $\theta \simless 88^o$, 
indicates that the total  mass loss rate in the wind
$\MDOT_w =2\times10^{-9}$~M$_{\odot}$~yr$^{-1}$, is dominated by 
the contribution from of the outer part of the disk.
We define $\MDOT_w$ as the cumulative mass loss rate for the region well
above the disk plane in which the flow gas passed the sonic point
(i.e., the angular integral is stopped early enough to avoid the exponential
density profile of the disk and any lower-velocity complex flow
component at $\theta$ near $90^o$).
The mass flux density has a prominent maximum at $\theta \sim 80^o$.

The column density in the wind increases gradually with $\theta$.
At $\theta\approx 87^o$, the base of the wind at $r=10^{10}$~cm,
the column density is $10^{23}~\rm cm^{-2}$. This column corresponds
to the highest column in the disk wind domain. For $\theta> 87^o$, the 
column density increases very sharply to $10^{29}~\rm cm^{-2}$ and it
corresponds to the column along the $\theta=90^o$ axis.

We checked that the line force is negligible everywhere in our
computational domain in this model. The line force is dynamically
unimportant in the region where the gas is not heated by the X-rays.
The density there is so high that most of the lines are optically thick
and the line force is smaller than the gas pressure gradient by 
at least one order of magnitude. We note that despite
the density being high in  the unheated gas, the line force there
can be higher than the radiation force due to electrons by a factor of 
$\sim10$.  
Although we do not resolve the transition region
between the regions of unheated and heated gas we do resolve the two regions. 
In particular, in the unheated gas  we resolve the transition between
the region where the total radiation force is  dominated by 
the radiation force due to 
electron scattering (at $\theta =90^o$) and the region where 
the total radiation force is dominated by the line force 
(at $\theta \sim 88^o$). 
We conclude then that in our model
H2 winds is thermally not line-driven. Resolving the chromosphere
will unlikely change the character of the wind
(see Section 5 for more discussion on this point).

The model~H2 shows that a LMXB with the high Eddington number, $\Gamma=0.67$,
can have a strong hot disk wind well within the Compton radius.
Two forces are driving the wind: the radiation pressure due to electron
scattering and thermal expansion. The line force is not dynamically
important because the strong X-ray radiation ionizes 
a relatively dense part of the disk. 
This result contrasts
with our results for AGN (see PSK) where we found that for a similar
Eddington number ($\Gamma=0.5$) and  the same
X-ray opacity, or even for the opacity  lower by as much as a factor of 100, 
the disk radiation can launch a wind from the disk in AGN
and the X-rays heat the line-driven wind not the disk.

\subsection{X-ray opacity effects}

To find out what  X-ray opacity is needed for the line force
to be dominant in  launching a wind we ran  models with:
$\kappa_{\rm X}=4\times 10^2~{\rm g^{-1}~cm ^2}$, 
$4\times10^3~{\rm g^{-1}~cm ^2}$, and $4\times10^4~{\rm g^{-1}~cm ^2}$
for $\xi \leq 10^5$. In all cases, 
we kept $\kappa_{\rm X}=0.4~{\rm g^{-1}~cm ^2}$ for $\xi > 10^5$.
We found that for $\kappa_{\rm X}= 4\times10^4~{\rm g^{-1}~cm ^2}$ (model H5), 
the X-rays can be sufficiently attenuated and the line force can launch and 
accelerate a wind to high velocities. 
We realize that X-ray opacity  we assumed in the model H5 is very high, 
unphysically so. Our detailed photoionization calculations using
the code XSTAR show that the maximum $\kappa_{\rm X}$ for $\xi < 10^5$
is $\sim 10^3$ (Proga \& Kallman 2001).
We therefore consider the model H5 only to check what 
the X-ray opacity would it take to sufficiently attenuate the X-rays  
for a LMXB. 

In the calculation of the model H5 we find (i) a hot, low density flow
in the polar region (ii) a dense, hot and fast wind, (iii) a dense, warm
and slow outflow from the disk and (iv) a warm flow, nearly
in the hydrostatic equilibrium, close to the disk
midplane.  Thus for such  high $\kappa_{\rm X}$, there is a new component
in the flow, the line-driven disk wind.
The model~H5 is an example where the X-ray radiation
heats the line-driven disk wind rather than the disk itself.
For $r = r_i$, at the base the line-driven wind, 
$\rho \sim 10^{-8}~\rm g~cm^{-3}$ and the gas temperature is $2 \times 10^5$~K.
The X-rays start to heat the line-driven wind, when the wind density
decreases below $\rho \sim 10^{-10}~\rm g~cm^{-3}$.

In the model H5, the lower part of the wind is driven 
by the line force (the maximum of $v_r(r_o)$ is
$\sim200$~km~$\rm s^{-1}$). Further from the disk midplane
the wind is heated by  X-rays, the line
force reduces to zero and the wind is 
thermally driven to high velocities
(the maximum of $v_r(r_o)$ is $ \sim2000$~km~$\rm s^{-1}$).
Thus the work of  the line force is taken over by the thermal expansion
where the wind is still moving relatively slowly and therefore the terminal
velocity of the wind is similar to the terminal velocity of 
a purely thermally driven wind. We note that the terminal
velocity of a purely line-driven wind (our model H0~PSD)
is $14000~\rm km~s^{-1}$ 
(see below for more details on this model). 
The total mass loss rate in the model H5
is $6\times10^{-11}$~M$_{\odot}$~yr$^{-1}$.
The column density in the hot wind is a few $10^{21}~\rm cm^{-2}$ 
whereas in the warm, line-driven outflow the column is
$\sim~10^{22}~\rm cm^{-2}$. The X-ray optical depth through the warm wind
is high ($\tau_{\rm X} > 1000$) because of the high $\kappa_{\rm X}$ for $\xi< 10^5$.

We calculated six models varying the X-ray opacity
between  $\kappa_{\rm X}=4\times10^{-1}~{\rm g^{-1}~cm ^2}$
and $\kappa_{\rm X}=4\times10^4~{\rm g^{-1}~cm ^2}$.
We note that the wind total mass loss rate, X-ray optical depth,
and column density increase with decreasing $\kappa_{\rm X}$, 
for $\kappa_{\rm X}<4\times10^4~{\rm g^{-1}~cm ^2}$ (see Table~1).
This increase of $\MDOT_w$, $\tau_{\rm X}$, and $N_H$ with
decreasing $\kappa_{\rm X}$
is simply related to the fact that the X-rays heat
denser disk regions for lower $\kappa_{\rm X}$.
We note that the column density for the model H1 is so high
that it may be opaque to the central
radiation (both the UV and the X-ray) at radii greater than $10^{10}$~cm.

In all models, except the one with the 
highest $\kappa_{\rm X}$ (i.e., except model H5), 
the thermal expansion dominates over the radiation
force in launching a wind. The radiation force, more precisely 
its radial component  due to electron scattering, 
is important in accelerating the wind to high velocities.

\subsection{X-ray heating effects}

For comparison, we calculated the PSD~99 type of model. In model H0~PSD, 
we  switched off the X-ray heating of the gas (see Table~1).
This corresponds to constant values for the parameters of the force
multiplier: $k=0.415$ and $M_{max}=2300$ (see PSK).
We assumed that the opacity for the central radiation is
$4\times10^{-1}~{\rm g^{-1}~cm ^2}$. The model H0~PSD shows that the 
line force could drive a disk wind if there were no X-ray heating.
Such a wind would be much faster than a thermally driven wind:
the terminal velocity of the model H0~PSD is 
$\sim 14,000~{\rm km~s^{-1}} \approx 3.2 v_{esc}(=\sqrt{GM_{NS}/r_i})$
close to that predicted by the theory of line-driven winds 
(e.g. CAK; PSD98; Proga 1999).
However the mass loss rate of the line-driven wind would be smaller
than for a thermally driven wind with relatively low
X-ray opacity (see Table ~1). 
We note that the line-driven wind has a steeper velocity gradient than 
the thermally driven wind;  the higher terminal velocity of the former
reflects that. Subsequently 
the density above the wind base is lower in the line-driven 
wind than in the thermally driven wind despite the drop of density
by $\sim 2$ orders of magnitude in the latter caused by the X-ray heating.

As a consistency check, we ran a test model with the line force switched
off and the  computational domain redefined to occupy the radial range
$r_i~=~10^2~r_\ast \leq r \leq \ r_o~=~ 10^3~r_\ast$.
All model parameters were as in the model H1. This test
showed that there is no wind coming from the lower radii.
For $r_i~<~10^3~r_\ast$, the X-rays produce an optically-thin isothermal corona
as BMS predicted. We conclude that our choice of the computational
domain, i.e., neglecting the  part of the disk with the temperature
$> 10^5$~K, does not change our results for the disk with
the temperature $\simless 10^5$~K.

\subsection{Dependence on $\Gamma$}

Figure~3 and Figure~4 shows the results for the model~L2 with 
$\MDOT_a=2\times10^{-9}$~M$_{\odot}$~yr$^{-1}$ ($\Gamma_D=0.067$). 
This model has
all parameters the same as the model~H2, except for the lower
mass loss accretion rate. The flow in this model is complex
and unsteady. We follow in this calculation (i) a hot, dense,
slow and complex corona and (ii) a warm zone near the disk midplane.
Although, the corona is hot it is non-isothermal contrary to the wind
in the models with the higher $\MDOT_{a}$ (i.e., higher $L_D$).

The velocity field in the corona is very complex and changes with time.
The maximum speed is $\sim 500$~km~$\rm s^{-1}$. The mass loss rate through the 
outer boundary is zero when the time averaged value is calculated.
In the models with $\MDOT_a=2\times10^{-8}$~M$_{\odot}$~yr$^{-1}$,
$\tau_{\rm X}$ and $N_H$ increase with decreasing $\kappa_{\rm X}$. In fact,
the optical depth and column density, 
for $\MDOT_a=2\times10^{-9}$~M$_{\odot}$~yr$^{-1}$
and $2\times10^{-8}$~M$_{\odot}$~yr$^{-1}$ are comparable for 
the same $\kappa_{\rm X}$.

The model~L2 illustrates the fact that the X-ray heating does not suffice
to drive a wind from 
the disk at radii between
$r_i~=~10^9~{\rm cm}$ and $r_o~=~10^{10}~{\rm cm}$. A high disk luminosity
is required so the radiation force due to electron scattering will
significantly lower gravity and allow thermal expansion to drive
a wind.

For comparison, we also calculated the PSD~99 type of model
for $\MDOT_a=2\times10^{-9}$~M$_{\odot}$~yr$^{-1}$ 
(model L0 PSD, see Table~1).
We found that for this lower disk luminosity,
the line force can not drive a wind even when the X-ray heating is
switched off. The line force produces a puffed-up disk instead.
The line force can launch material from the disk but fails
to accelerate it to the escape velocity. The reason for the latter
is that the luminosity of the UV emitting disk is highly sub-Eddington
(i.e.,  $\Gamma_D({\rm UV})\equiv L_D(UV)/L_{Edd} \simless 10^{-3}$, 
see Appendix for details how estimate  $\Gamma_D({\rm UV}$)).
Then the line force can not exceed gravity even when the force multiplier
reaches its maximum value of $\sim~2000$. Therefore   
a relatively luminous accretion disk 
with  $\Gamma_D=0.067$ ($\MDOT_a=2\times10^{-9}$~M$_{\odot}$~yr$^{-1}$ )
has too little luminosity in the UV  and cannot produce
a line-driven disk because it fails to satisfy the basic requirement
for the strength of the UV radiation: 
\begin{equation}
\Gamma_D({\rm UV})[1+ M_{max}(\xi)] > 1
\end{equation}
(e.g., Proga 1999). 
The above equation is a modified  version of equation 6 in Proga (1999),
who consider systems with most of their luminosity in the UV 
($\Gamma_D \sim \Gamma_D(UV)$) and 
not contribution from X-rays ($M_{max}=constant$). 
For simplicity, we set the so-called finite 
disk correction factor to 1 in Proga's eq. 6. 
Here we emphasize that the Eddington number of the UV zone matters
more than the Eddington number of the whole disk 
and that  the maximum force multiplier is a function
of the photoionization parameter with the following limiting
behavior: 
\begin{eqnarray}
\lim_{\xi \rightarrow \infty}~M_{max} & \approx & 0 \\
\lim_{\xi \rightarrow 0}~M_{max} & \approx & 2000.
\end{eqnarray}

We finish this section with the note that
in all our models, we took into account
the contribution from the irradiation to the disk intensity (see Section~3)
assuming that no irradiation is lost between the source and the disk
surface. Our finding that the disk wind or corona are optically thin
provides a justification to this assumption.

\section{Discussion}

We have calculated a series of models for winds from accretion disks in 
LMXB. Our calculations allow for self-consistent determination
whether the disk wind is driven by thermal expansion or radiation
pressure. The calculations show that the UV emitting accretion
disk cannot produce a wind that could shield itself from the 
strong ionization radiation emitted from the central object.
This inhibition of the line-driven
wind indicates that irradiation penetrates below the so-called
wind critical point where, in the CAK type models, the wind
mass loss rate is determined.
A disk wind from the UV emitting disk is only possible 
for very luminous sources whose luminosity is within a factor $\sim 2$
of the Eddington limit. 
However in such a case,
the disk wind is very hot and driven by thermal expansion assisted by 
the radiation pressure due to electron scattering. 
A cooler, line-driven wind could only be  possible for an unphysically
high X-ray opacity. Our results are consistent with the UV observations
of LMXBs which show no spectral features associated with a strong
and fast disk wind, although they fail to account for the single peaked
UV emission lines observed from some objects.

We model here a wind from a disk that is Keplerian,
geometrically-thin and optically thick. 
To account for the disk irradiation by the central
object, we consider the disk photosphere with the concave  shape
($h \propto r ^{9/7}$). The shape of the disk is important
because it can significantly change the role of X-ray irradiation.
Dubus et al. (1999) showed that a self-consistent disk model produces 
a convex disk. Thus the outer disk regions can be screened from X-ray 
irradiation.
Observations show however that outer regions of accretion disks in LMXBs are
irradiated, so they are intercepting X-rays. Dubus et al. concluded
that these disks could  be non-planar or that the irradiating 
source could be outside the equatorial plane and/or not point-like or both.
Unless the inner disk screens the UV emitting disk and allows line driving, 
we anticipate that the convex disk would not change our main result~--~
X-rays will still inhibit line driving. 

Our calculations for plausible $\kappa_{\rm X}$ show that there is a very 
sharp transition between the disk and the X-ray heated corona or wind. 
For example, the gas temperature increases by $\sim 2$ orders of magnitude 
within one grid zone in the model H2. This dramatic change in the gas
properties is partially related to an inadequate resolution on our 
relatively large spatial grid and to our simplistic treatment of 
the photoionization structure. 
In particular, we do not consider the optical depth
effects while calculating the line cooling. More realistic treatment of 
photoionization could affect the transition zone between warm and very hot 
gas (Raymond 1993; Ko \& Kallman 1994). We note that the transition zone is 
typically very thin and hard to resolve even in time independent
calculations (e.g., Raymond 1993; Ko \& Kallman 1994). In previous 
time-dependent calculations, simulations start where the gas is already
X-ray heated (i.e., Woods et al. 1996).
Generally, we expect that the presence 
of a  zone with intermediate temperature and density, the chromosphere,
may somewhat change dynamics of the wind but only near its base. For example,
the chromosphere may provide more shielding to X-rays and this
may help to permit launching of a line-driven wind from the disk so the 
mass loss rate will be determined by the line force. 
Subsequently, X-rays
will irradiate that wind instead of the disk. 
However, we predict that 
the line-driven wind could not be accelerated to very high velocities
because the irradiation will fully ionize the wind very close
to the disk.

In our model, we do not take into account the effects of 
scattering of photons in a wind or corona or both 
(e.g., Ostriker, McKee \& Klein 1991; Murray et al. 1994). 
Ostriker et al. (1991) considered systems in which the scattered radiation
was assumed to be dominated by singly scattered photons.
While this assumption limited their models to systems 
with $L \simless 0.13 (T_{IC}/10^8~K)^{-1.4} L_{Edd}$ (eq. 64 in Ostriker 
et al), they were
able to treat the radiative transfer accurately in two dimensions.
The assumption of hydrostatic equilibrium limited their models
to the inner part of the disk ($r/R_{IC} \simless 0.1$).
Using above conditions, we find that the single scattering limit
may apply to our models because we consider systems with luminosity
below  $0.13 (T_{IC}/10^8~K)^{-1.4} L_{Edd}\approx 0.9$ and
our computational domain is for $r/R_{IC} \simless 0.1$.
Therefore we can use Ostriker et al.'s expressions to estimate
the importance of the scattered radiation as compared to the direct
radiation. Using their equations 73 and 75, we found that the scattered
radiation is comparable to the direct one at our inner boundary.
We reached the same conclusion using Murray et al.'s (1994) results.
Murray et al. (1994) made the assumptions similar to those of Ostriker et al. 
with a difference in the treatment of the radiation field. Murray et al.'s
method is more approximate, the scattered radiation is approximated
by flux-limited diffusion, but includes multiply scattered radiation.
The parameters of Murray et al.'s model 5 (i.e., $M_{NS}=1 \MSUN$
and $T_{IC}=0.1 \times 10^8$~K) are the closest to our model
parameters. In particular, Figure~8 in Murray et al., shows that for 
$L/L_{Edd}=0.590$
the scattered radiation is comparable to the direct one at $r/R_{IC}=0.01$
(our inner radius) and decreases with radius whereas the direct
radiation stays constant. Thus the scattered radiation is important
for some radii. However we anticipate that the inclusion of the effects of 
scattered radiation would not change
qualitatively our results because
the scattering of  X-rays in the wind or corona would most
likely increase the irradiation and suppress the line driving 
even more compared to our calculations. Our results give
an lower limit on mass loss rate of thermally driven winds.
We plan future work on the effects of scattered radiation and to
resolve the chromosphere. We note that both need to be done at the same
time because the scattered radiation may affect the
properties of the chromosphere.

Having said that line driving is not important in our calculations
of X-ray heated disks in LMXB we can ask if our calculations are
consistent with previous calculations, in particular, 
with those by Woods et al. (1996) who extended and improved the analytic
predictions by BMS.  As Woods et al., 
we included non-Compton processes in radiative heating and cooling
(i.e., photoionization heating, and line and bremsstrahlung cooling,
Woods at al. did not include the latter)
Also as Woods et al., we  did not 
include the effects of scattering of photons. 
However contrary to us,
they did not include the radiation force and attenuation of 
the X-rays.  
Our calculations also differ
in the location of the low boundary, the base of the wind, and in
the conditions along the boundary.
In particular, Woods et al. obtained the gas temperature along the base 
by solving the thermal equilibrium equations with a given, 
radius-independent photoionization parameter whereas we assume
that the gas temperature is the same as 
the effective temperature of the optically thick accretion disk 
(Shakura \& Sunyaev 1973).
Woods et al. obtained the density from the equation of state whereas
we assume a fixed density.
Generally, Woods et al.'s lower boundary conditions are specified for 
the gas which is in the part of the disk atmosphere heated by X-rays 
whereas our lower 
boundary conditions are for the gas in the disk
photosphere where X-rays do not penetrate. This difference in the boundary
conditions is also reflected in the location of the boundary: we adopted
the offset of the $\theta=90^o$ axis
from the disk midplane of $3\Gamma_D r_\ast$ whereas Woods et al.
adopted the offset of $\simgreat 100 r_\ast$ (see their eq. 3.1).
We note that the  lower boundary adopted by Woods et al. can not be
applied to the case where the gas is optically thick to X-rays 
(the case we explore in this paper). 
In such a case, we can not assume a given photoionization parameter
along the boundary because the photoionization parameter is 
a function of the optical depth through the wind the structure of 
which we want to calculate.

To make a comparison with models presented by Woods et al. (1996),  we 
ran several tests
with the radiation force and 
X-ray attenuation switched off. Specifically, 
we ran several tests 
for $\kappa_x=0$ and $10^9$~cm $\ge r \ge 10^{11}  $~cm using ours and
Woods at el.'s lower boundary conditions. We found that 
models with our and Woods et.~al.'s  boundary conditions are similar although
models with our boundary conditions settle down into a steady state
more slowly. 

Figure~5 shows a comparison between our results with and without
radiation force and Woods et al.'s results. The figure shows
the mass flux density as a function of radius for various models.
The solid line represents our 
results for $\Gamma_D=0.667$ with radiation force and X-ray attenuation 
switched off whereas the dashed line represents our results with radiation 
force switched on. In both tests, we assumed  $\kappa_X=\kappa_{UV}=0$. 
For comparison,
the figure  also shows the analytic fit (the triple-dot dashed line) 
by Woods at al. 1995 to their numerical results (see their eq. 5.2).
For the constant $C_0$, in Woods et al.'s expression, we adopted 
$10^{-4}$.

Our tests show that our model reproduces the Woods et al (1996) results
when we adopt the same assumptions. In particular,
we found 
that the mass flux density scales like $r^{-2}$ 
for large radii and have a maximum at $\sim 0.2 R_{IC}$.
However when we add the radiation force due to electron scattering, 
the $r^{-2}$ scaling extends for smaller radii and the maximum
of the mass flux density occurs at $\sim 0.01 R_{IC}$.

We conclude that our results for a X-ray heated disk are consistent with
previous models (e.g., BMS, Ostriker et al 1991, Woods et al. 1996).
In particular, we showed that the solution depends on the location of the disk 
surface with respect to the Compton radius, $R_{IC}$, and the disk luminosity 
in units of the Eddington limit. A new element in our model, the inclusion of 
the radiation force, changes the solutions for LMXBs for systems with high 
luminosities (e.g. $\Gamma_D \simgreat 0.7$) by producing a hot robust disk 
wind well inside the Compton radius (at the distance from the center smaller 
than the Compton radius by a factor of $\sim 100$). This inner disk wind, 
for a very luminous system, is due to lowering of the effective gravity by 
the radiation force due to electrons and subsequently lowering the escape 
velocity. We illustrate the latter by considering the 
binding energy of a test particle under the influence of  the gravity,
the radiation force due to electrons  and the centrifugal
force.  We assume that  the particle is initially
on  the Keplerian orbit where the centrifugal force cancels out the gravity.
These conditions are as those for the lower boundary  of 
our hydrodynamical calculations. 
The radiation force is a complex function of the position in the wind.
In particular, the force due a disk must be evaluated numerically
(e.g., Icke 1981; PSD~98). The force due a central object is much
less complex especially for large radii.
For our purposes here, we will consider explicitly
only the force due to the central object as it can be approximated
by a simple analytic formula.
For large radii, the radiation force to due the 
central object is purely radiation and can be written as
\begin{equation}
F^{rad, e}_\ast= \frac{ G M_{NS}}{r^2} \left( \frac{\Gamma_\ast}{2}
+\frac{\Gamma_\ast}{2} \cos{\theta}\right).
\end{equation}
Although the force due to the central object is radial it depends on
$\theta$ because the presence of the optically thick disk
which obscures half of the object. The disk obscuration
is the strongest at $\theta=90^o$ where only half of the object
is visible and therefore 
$F^{rad, e}_\ast= \frac{ G M_{NS}}{r^2} \frac{\Gamma_\ast}{2}$. 
At $\theta=0^o$, the disk does not obscure the object
at all and therefore 
$F^{rad, e}_\ast= \frac{ G M_{NS}}{r^2} \Gamma_\ast$.

The particular dependence of the radiation force on the position
prevents us from  expressing the force as a gradient of a potential.
Therefore we have to assume certain geometry for the trajectory.  
As an analytically tractable case,
we assume the helical trajectory that is contained within a straight cone.
In cylindrical coordinates $(R,Z)$, the position of the particle
along the trajectory can be expressed as: $R=R_f + l \cos{i}$
and $Z= l \sin{i}$, where $i$ is the angle between the trajectory cone
and the disk midplane, $R_f$ is the footpoint where the particle leaves
the disk and $l$ is the distance to the footpoint.
The total force along the trajectory
acting on the particle in a rotating frame is:
\begin{equation}
F^{tot}_l= \frac{G M_{NS}(l+R_f\cos(i)) }{(R_f^2+l^2+2l\cos(i))^{3/2}} 
\left[-1
+\frac{\Gamma}{2}\left(1 + \frac{l\sin{i}}{(R_f^2+l^2+2l\cos(i))^{1/2}}\right)\right]
+\frac{G M_{NS}R_f\cos(i)}{(R_f+l\cos(i))^3} + F^{rad, e}_{D,l}.
\end{equation}
The first term in eq. 20 corresponds the gravity reduced by the
radiation force due to the central object, the
second term corresponds to the centrifugal force while the third
term corresponds to the radiation force due to the disk.
The latter needs to be determined numerically and we leave it here
in a general form.

To calculate the energy needed for a particle to escape
from the disk (i.e., the binding energy), we 
integrate the total force along the trajectory:
\begin{equation}
E \equiv \int^{0}_{\infty} F^{tot}_l dl= \frac{G M_{NS}}{R_f}
\left[1-\frac{\Gamma}{2} \left(1 + \frac{i}{2}\right)\right] 
-\frac{G M_{NS}}{2R_f}
+\int^{0}_{\infty} F^{rad,e}_{D,l} dl.
\end{equation}
The first term in eq. 21 is the gravitational potential energy
corrected for the radiation force due to the central object. 
The second term is the kinetic energy of the rotational motion
and the third term is negative and is the work done by the radiation force
from the disk.
We assume an optically disk along the lower boundary.
Therefore along the lower boundary the  radiation force  is zero,
and the centrifugal forces balances the gravity.
However above the lower boundary, gas quickly becomes optically
thin and the radiation force is non zero. Consequently, 
the total force in the radial direction is positive near the lower boundary.

The classical Compton radius is defined 
as the radius on the Keplerian disk at which the 
escape velocity equals to the gas thermal velocity.
Using this definition and eq. 21,  we can calculate 
the Compton radius corrected for the
effects of the radiation force, $\bar{R}_{IC}$ as
\begin{equation}
\bar{R}_{IC}= R_{IC} [1-\Gamma_\ast(1+i/2)
+\frac{2R_f}{G M_{NS}}\int^{0}_{\infty} F^{rad,e}_{D,l} dl].
\end{equation}
We note that $\bar{R}_{IC}$ depends
on $\bar{R}_{IC}$ through the disk term.
However for our illustrative purposes let us assume for a moment that the
disk contribution is negligible.
For $\Gamma=0.7$ and $i=45^o$ (note that $i$
is in radians in eq. 21-24), the correction for the radiation force, 
is $\sim 0.07$. Numerical simulations of disk winds driven only by
thermal expansion  showed that the maximum of 
the mass flux density is at $\sim 0.2 R_{IC}$ 
(see Woods et al. 1996 and also above).
In our hydrodynamical calculations of disk winds driven by thermal expansion
and radiation force, we found that the maximum of 
the mass flux density is at $\sim 0.01 R_{IC}$.
If we correct for the radiation force the Compton radius computed 
by Woods et al. we  find 0.014 $R_{IC}$ which is in an agreement 
with our numerical calculations.

We note that the Compton radius depends on the luminosity of the central 
object and disk but also on the geometry of the streamlines relative
to the geometry of the radiation field (e.g., 
$\bar{R}_{IC}$ depends the angle, $i$, when the streamlines are straight). 
Generally, the inclination angle between the poloidal velocity
and the disk changes with the position. 
The radiation force vector from the disk is a complex  
function of the position that can be evaluated only numerically
(e.g., Icke 1981, PSD~98). Therefore it is not straightforward to incorporate
this force to our analysis presented above. However there are some 
characteristics of the disk radiation that can help us to understand 
the geometry of the disk radiation field and understand why the disk
contribution does not  change the Compton radius
even when the disk luminosity equals to the central object luminosity
as in our hydrodynamical calculations.
In particular,
for  large radii, the radial component of the disk radiation
scales as $\cos(\theta)/r^2$ while the $\theta$
component scales as    $1/r^3$ for $\theta\sim 90^0$.
This just means that near the disk, the radiation from the disk
is normal to the disk.
Then the escape velocity is most reduced
for the streamlines normal to the disk and least reduced 
for the streamlines tangent to the disk. However
the geometry of the streamlines is such that
the gas does not gain much of support from the disk radiation
because the streamlines are not aligned with the radiation flux
vector from the disk. The streamlines are rather aligned
with the total radiation flux. The poloidal velocity  (Fig. 1d)  shows 
that the gas streamlines are 
predominantly tangent to the disk for most of their length
and therefore the escape velocity is little effected by the radiation from
the disk.

We also note that the correction of the Compton radius for the radiation
force depends on the conditions assumed along the lower boundary.
For example, if we assume that gas along  the  lower boundary is optically
thin (e.g., BMS and Woods et. al. 1996) then radial 
momentum balance requires the centrifugal force to cancel out
the gravity  minus  radiation force. Consequently, the test particle
is not on the Keplerian orbit and the rotational kinetic
energy is reduce by a factor of ($1-\Gamma_\ast/2$)
compared to the optically thick case we considered in above.
Then for the optically thin gas along the lower boundary, 
the binding energy is 
\begin{equation}
E \equiv \int^{0}_{\infty} F^{tot}_l dl= \frac{G M_{NS}}{R_f}
\left[1-\frac{\Gamma}{2} \left(1 + \frac{i}{2}\right)\right] 
-\frac{G M_{NS}}{2R_f}\left(1-\frac{\Gamma_\ast}{2}\right)
+\int^{0}_{\infty} F^{rad,e}_{D,l} dl.
\end{equation}
Then the only difference between eq. 23 and eq 21 is in the term
corresponding to the  rotational kinetic energy 
[the factor of ($1-\Gamma_\ast/2)$ in eq. 23].
From the definition and eq. 23, the Compton radius corrected 
for the effects of the radiation force can be written as
\begin{equation}
\bar{R}_{IC}= R_{IC} [1-\Gamma_\ast(1+i)/2
+\frac{2R_f}{G M_{NS}}\int^{0}_{\infty} F^{rad,e}_{D,l} dl].
\end{equation}
Thus for the optically thin case, 
the  correction for the radiation force is smaller than in the
optically thick case. The reason for this difference 
is simply due to
the fact that at the lower boundary the rotational kinetic energy 
is smaller and therefore the gravitational binding is stronger
in the optically thin case than in the optically thick case.

An intriguing aspect of our calculations is that 
results for LMXB differ dramatically from the results for AGN.
As we mentioned in Section~1, AGN resemble LMXBs
in their X-ray/bolometric ratios and recent calculations  by PSK
demonstrated that AGN disk winds can provide sufficient shielding
against X-ray ionization to allow significant mass loss from the disk.
In the case of LMXBs, the shielding from strong X-rays is insufficient.
Perhaps the simplest way to try to  explain this difference is to compare
the Compton radius, $R_{IC}$  with the minimum radius at which the disk
emits UV capable of efficiently driving a wind, e.g., 
$r_{\rm UV}=r_D(T_D=50 000~\rm K$). 
For LMXBs, the ratio between these two radii, $r_{UV}/R_{IC}$ is 
$\sim 5\times 10^{-2}$
whereas for AGN is $\sim 10^{-4}$. Thus in the UV emitting disk,
the X-ray heating is dynamically much more
significant for LMXB than for AGN.

In addition to the fact that X-ray heating is dynamically {\it more} 
important in LMXBs than in AGN, we can show that 
line driving is dynamically {\it less} important in LMXBs than in AGN. 
To address this question in a quantitative way,
we derive analytic formulae to estimate the photoionization
parameter and the optical depth in a radiation-driven disk wind
(see Appendix). Our approach is simplified in that it assumes
that the wind is in a steady state and is limited to the region
near the base of the wind. As we discussed in Section~4, one of the key 
parameters determining the importance of UV line driving, 
is the Eddington number of the UV zone, $\Gamma_D({\rm UV})$ that depends
on radius for a given total Eddington number (see Appendix). 
The limiting behavior of $\Gamma_D({\rm UV})$ as a function of $r$,
i.e., $\Gamma_D({\rm UV})=\Gamma_D$ at $r=r_\ast$ 
motivates us to use the radius
in units of the inner radius of the accretion disk, $r'\equiv r/r_\ast$ 
to describe our results for line-driven disk winds. 

Equations A13 and A18 provide estimates for the photoionization parameter
and X-ray optical depth  at the base of the wind, respectively.
Equation A14 shows that the photoionization
parameter for an optically thin case, $\xi_{\tau_{\rm X}=0}$
decreases  with decreasing $r'$.
($\xi_{\tau_{\rm X}=0} \propto r'^{1.67}$ for $\alpha=0.6$).
Therefore, in the optically thin case, 
the line-driven disk wind is more likely  to 
avoid strong ionization at small radii than at large radii.
However even though this effect gives the needed trend,
it is not sufficient to  reduce the photoionization
parameter so the line force can operate in the case of LMXBs.
The photoionization parameter can be sufficiently reduced (i.e., $\xi <
10^2$)  if X-ray attenuation is taken into account.
Equation A18 shows that $\tau_{\rm X}$ per unit length increases
with decreasing radius ($\tau_{\rm X} \propto r'^{-2.67}$ for $\alpha=0.6$).
This implies that a line-driven wind can better shield itself
at smaller radii than at larger.
To illustrate the difference it makes if the wind
is launched from different radii we estimate the photoionization
parameter in the optically thick case 
$\exp(-\tau_{\rm X})~\xi_{\tau_{\rm X}=0} $, where the first factor 
is given by eq. A17 and the second is given by eq. A13.
Figure~6 shows the photoionization parameter,  as a function
of $r'$ for various $\kappa_{\rm X}$. For the wind velocity, $v$ 
we adopted the value of  $20$~km~$s^{-1}$ while for all other parameters 
we adopted values  typical for a LMXB (see Section 3) which, 
as we already mentioned, are very similar to those
for AGN. As noted above, the only difference  between LMXB and AGN 
is the radius where the disk is emitting the UV radiation.
Figure~6 illustrates the fact that $\xi$ is very sensitive
to $r'$ and $\kappa_{X}$. This is in qualitative agreement with our
detailed
numerical calculations.  However Figure~6 shows  even  a good quantitative
agreement between our analytic and numerical results: the line-driven disk
wind can shield itself at $r'=10^3$ 
for $\kappa_{\rm X} \simgreat 4\times10^{-1}~{\rm g^{-1}~cm ^2}$, as in
AGN,
but at $r'=10^4$ a much higher opacity would be required
($\kappa_{\rm X} \simgreat 4\times10^{4}~{\rm g^{-1}~cm ^2}$) to
allow strong line driving in LMXB.

\section{Conclusions}

We have studied radiation- and thermally driven winds from luminous
accretion disks irradiated by strong X-rays in LMXBs. We have used
numerical methods to solve the two-dimensional, time-dependent
equations of hydrodynamics. We have accounted for the radiation
force due to electron scattering and spectral lines. We have estimated
the latter using a generalized multidimensional formulation of the Sobolev
approximation. As for thermal driving, we have taken into account
the effects of the strong central radiation on the disk and wind
photoionization structure and thermodynamics.

We find that the local disk radiation cannot launch a wind
from the disk because of strong ionizing radiation from 
the central object. Unphysically large X-ray opacities would
be required to shield the UV emitting disk and allow
the line force to drive a disk wind.
However the same X-ray radiation that inhibits  line driving
can heat the disk and produce a hot wind or corona above the disk.
In this respect our calculations are consistent with past work
on the dynamics of coronae and winds from accretion disks in LMXBs.
The main difference between our  work and previous work 
is related to the fact that we included the radiation force:
although the line  force is dynamically unimportant, the radiation 
force due to electrons can be important for very luminous
sources. We found  that the reduction of the effective gravity
caused  by the radiation force due to electrons can expands
inwards the radial range for the thermally driven disk wind.

The application of our model to LMXBs provides a test
for our self-consistent treatment of the X-ray heating and line-driving.
Our results are consistent with the UV observations
of LMXBs which show no obvious spectral features associated with a strong
and fast disk wind. As predicted by past work, the thermally driven disk 
winds and coronae are optically thin and are not likely to modify the 
UV spectrum of the disk photosphere and chromosphere.

Our calculations do not fully resolve the transition between
the X-ray heated gas and the unheated gas.
It is possible that
the chromosphere may provide more shielding to X-rays 
than our calculations produce and this
may help to permit launching of a line-driven wind from the disk.  
Subsequently, X-rays will irradiate that wind instead of the disk. 
Generally, we expect that the presence 
of the chromosphere may change the wind mass loss rate
but not the wind velocity because its influence is greatest
very close to the disk photosphere.

To put our present calculations in a larger perspective, we 
compared our results for LMXBs with those for AGN by PSK.
AGN resemble LMXBs in their X-ray/bolometric ratios,
yet they significantly differ, for example, in their UV radiation.
Some AGN, in particular QSOs, show UV spectral features that 
indicate powerful winds: broad absorption lines.
Recent calculations by PSK demonstrated that  
line-driven disk winds can provide sufficient shielding
against X-ray ionization to allow significant mass loss from the disk.
Here we have shown why we do not see such disk winds in LMXBs: 
the shielding from strong X-rays is insufficient.
To be able to  assess the impact of X-ray heating
upon driving of a disk wind by the line force in any system
with an accretion disk we derived  analytic formulae.
The key parameter determining the role of the line force is 
not merely the presence of the luminous UV zone in the disk
and of the X-rays
but the distance of this UV zone from the center. 
In other words, the Eddington number of the UV zone, 
$\Gamma_D({\rm UV})\equiv L_D(UV)/L_{Edd}$. 
Based on our results and results from PSD98 and Proga (1999),
we  can write the  condition for a radiation-driven wind 
$\Gamma_D({\rm UV})[1+ M_{max}(\xi)] > 1$ (see Section~4 and
eq. 6 in Proga 1999).
Using the parameters from here and PSK, we find
that the above condition can be more likely satisfied  
in the AGN case where $\Gamma_D({\rm UV})\approx 0.1$
than in the LMXB case where $\Gamma_D({\rm UV})\simless 0.001$. Generally,
the closer the UV zone to the center (higher $\Gamma_{UV}$), the stronger
the line force and subsequently the denser line-driven disk wind. 
The density of the disk wind critically determines
whether the wind will stay in a lower ionization state in the presence
of the X-ray radiation (whether $\xi$ will be low) and be further accelerated 
by the line force to supersonic velocities (whether $M_{max}$ will be high).

ACKNOWLEDGEMENTS: 
We thank Scott Kenyon, John Raymond and James Stone
for comments on an early draft of this paper.
We also thank an anonymous referee for comments
that helped us improve our presentation.
This work was performed while DP held
a National Research Council Research Associateship at NASA/GSFC.
Computations were mainly supported by NASA grant NRA-97-12-055-154.
Some computations were also performed at Imperial College Parallel Computing
Center.

\begin{table*}
\footnotesize
\begin{center}
\caption{ Summary of parameter survey.}
\begin{tabular}{l c  c c c c c  l  } \\ \hline 
         &       &    &            &                       &   & &   \\
Run & $\MDOT_a$  & $\kappa_{\rm X}$ & $\MDOT_w$   &  $v_r$  & $\tau_{\rm X}$ & $N_H$  & comments  \\ 
 &  (M$_{\odot}$ yr$^{-1}$) &  (${\rm 0.4 g^{-1}~cm ^2}$) & (M$_{\odot}$ yr$^{-1}$) & $(\rm km~s^{-1})$ &   & ($\rm cm^{-2}$) &   \\ \hline  

         &                &   &                        & & & &   \\
L0     &  $ 2\times10^{-9}$ & $10^0$  &   &    & $6\times10^{-1}$ & $8\times10^{23}$ &    non-iso. complex corona \\
L1     &  $ 2\times10^{-9}$ & $10^1$  &   &     & $1\times10^0$ & $2\times10^{23}$ &    non-iso. complex corona \\
L2     &  $ 2\times10^{-9}$ & $10^2$  &   &     & $2\times10^{-2}$ & 
$3\times10^{22}$ &    non-iso. complex corona \\
         &                &   &                        & & & &   \\
L0~PSD     &  $ 2\times10^{-9}$ & $10^0$  &   &    & $6\times10^{-3}$ & $8\times10^{21}$ &    puffed up disk \\

 &         &                &   &            &  & &       \\
H0     &  $2\times10^{-8}$ & $10^{0}$  & $1\times10^{-8}$  &   2000  & $7\times10^{-1}$ &$1\times10^{24}$ &   iso. wind \\
H1     &  $2\times10^{-8}$ & $10^{1}$  & $6\times10^{-9}$  &   1900  & $2\times10^{-1}$ &$3\times10^{23}$ &   iso. wind \\
H2     &  $2\times10^{-8}$ & $10^2$  & $2\times10^{-9}$  &   2000  & $5\times10^{-2}$ &$9\times10^{22}$ &   iso. wind \\
H3     &  $2\times10^{-8}$ & $10^3$  & $6\times10^{-10}$  &   1900  & $2\times10^{-2}$ &$3\times10^{22}$ &   iso. wind \\
H4     &  $2\times10^{-8}$ & $10^4$  & $1\times10^{-10}$  &   2000  & $4\times10^{-3}$ &$6\times10^{21}$ &   iso. wind \\
H5     &  $2\times10^{-8}$ & $10^5$  & $6\times10^{-11}$  &   2000  & 
$1\times10^{3}$ &$2\times10^{22}$ &    iso. wind and line-driven wind \\

 &         &                &   &            &  & &       \\

H0~PSD     &  $2\times10^{-8}$ & $10^0$  & $3\times10^{-10}$  &   14000  & 
$1\times10^{0}$ &$2\times10^{24}$ &   line-driven wind \\

 &         &                &   &            &  & &       \\

\hline
\end{tabular}

\end{center}
\normalsize
\end{table*}

\onecolumn
\appendix

\section{Derivation of an analytic formula for the photoionization parameter
at the base of a radiation-driven disk wind}

The photoionization parameter for an optically thin case 
and a point-like X-ray source can be written
as:
\begin{equation}
\xi_{\tau_{\rm X}=0} = \frac{ L_{\rm X}}{n r^2},
\end{equation}
where $n$ is the number density of the gas ($={\rho}/({m_p \mu })$ , where  
$m_p$ is the proton mass, and $\mu$ is the mean molecular weight).
Using our parameterization (see Section~3 and PSD~98) and the expression 
of the accretion disk luminosity (e.g. Pringle 1981), 
the X-ray luminosity is
\begin{equation}
\xi_{\tau_{\rm X}=0}= x f_{\rm X}\frac{ L_{\rm Edd} \Gamma_D}{n r^2},
\end{equation}
where $L_{\rm Edd}=\frac{ 4 \pi c G M_{\ast}}{\sigma_e}$ 
is the Eddington limit, and $M_{\ast}$ is the mass of the accreting
central object. 

To calculate the photoionization parameter, we 
need to know the number density in the wind.
We estimate this density using the results from PSD~98 and Proga~(1999).
Proga(1999) found that stellar and disk wind driven by radiation,
within CAK framework, are very similar as far as mass-loss rates are
concerned. In particular, we can use  analytic results for stellar
wind to rescale, in a first-order approximation, results for disk winds.
We follow this and approximate the mass loss rate for the disk wind, 
$\MDOT_W$ as
\begin{equation}
\MDOT_W \approx  \frac{L_{\rm Edd}}{c v_{th}}
\frac{\alpha}{1-\alpha}(1-\Gamma_D)^{-(1-\alpha)/\alpha}
\left[ (1-\alpha) k\Gamma_D \right]^{1/\alpha}
\end{equation}
(e.g., CAK, Proga 1999).

Our next step in estimating the wind density is to assume that
the mass loss rate from a given annulus on the disk surface, scales with the 
luminosity of the annulus, in units of the Eddington  luminosity, 
in  the same manner as $\MDOT_W$ scales with
$\Gamma_D$. For example, if we consider an annulus on the non-irradiated
disk  (i.e., $I_{D, total}=I_{D}$, see Section~3)
between the two radii, $r_1$ and $r_2$ then the Eddington number
for this annulus is:
\begin{equation}
\Gamma_D(r'_1, r'_2)= \frac{4 \pi {r_{\ast}}^2}{L_{Edd}} \int_{r'_1}^{r'_2} 
r'_D I_{D, total}(r'_D) dr'_D=3 \Gamma_D (f(r'_1)-f(r'_2))
\end{equation}
where 
$r'_1=r_1/r_{\ast}$, $r'_D=r_D/r_{\ast}$, and 
$f(r'_1)={r'_1}^{-1}-\frac{2}{3} {r'_1}^{-3/2}$ with $r_\ast$ being
the radius of the central object.
Our assumption of the non-irradiated disk, is here  justified because we 
use the  results  from PSD~98 and Proga's (1999), obtained for a flat disk
where the contribution to the disk intensity, from the irradiation
is negligible for large $r_D$ and $x\sim 1$.
We estimate the Eddington number for the UV emitting disk as
\begin{equation}
\Gamma_D({\rm UV})\simeq  \Gamma_D(r_u, r_l), 
\end{equation}
where $r_u$ and $r_l$ are the radius at which the disk temperature
is $50 000$~K and $10 000$~K, respectively.
Equation A4 for $r_2= \infty$ gives the partial disk Eddington number:
\begin{equation}
\Gamma_D(r_1)\equiv\Gamma_D(r_1, \infty)= 3 \Gamma f(r'_1).
\end{equation}
Then we can write that the mass loss rate from 
the disk between $r'_1$ and $\infty$ is
\begin{equation}
\MDOT_W(r'_1) \approx  \frac{L_{\rm Edd}}{c v_{th}}
\frac{\alpha}{1-\alpha}[1-3\Gamma_D f(r'_1)]^{-(1-\alpha)/\alpha}
\left[ (1-\alpha) k 3 \Gamma_D f(r'_1) \right]^{1/\alpha}.
\end{equation}
On the other hand, the mass loss rate from the disk surface
can be written as
\begin{equation}
\MDOT_w(r'_1)= 4 \pi {r_{\ast}}^2 \int_{r'_1}^{\infty} r'_D \rho v dr'_D
\end{equation}
where $\rho v$ is the mass flux density from the disk surface.
The mass flux density from eq. A8 is
\begin{equation}
(\rho v)(r'_1)= -\frac{1}{4\pi {r_{\ast}}^2}\frac{1}{r'_1}\frac{d \MDOT_W(r'_1)}{d r'_1}.
\end{equation}
Combining eqs A7 and A9 yields:
\begin{equation}
(\rho v)(r'_1)\approx-\frac{1}{4 \pi r^2_{\ast}}
\frac{1}{r'_1}\frac{L_{\rm Edd}}{c v_{th}}
\frac{1}{(1-\alpha)}
\left[ 3 (1-\alpha) k\Gamma_D \right]^{1/\alpha} f'(r'_1)
f(r'_1)^\frac{1-\alpha}{\alpha},
\end{equation}
where $f'(r'_1)=df(r'_1)/dr'_1$. We derived eq. A10, in the limit: for $r'_1>1$,
where $1-\Gamma_D(r'_1)\approx 1$  and $3 (1-\alpha) \Gamma_D << f(r'_1)^{-1}$.
We note that in this regime:
\begin{equation}
-f'(r'_1)f(r'_1)^\frac{1-\alpha}{\alpha} = {r'_1}^{-2} 
{r'_1}^{-\frac{1-\alpha}{\alpha}}
\end{equation}

We consider the regions close to the disk photosphere, where the disk wind
is moving in the direction perpendicular to the disk midplain (i.e.,
the poloidal velocity $v$ has only non-zero component in the direction
perpendicular to the disk).
In such a case, the continue  equation and 
eq. A10 allow us to estimate the  density as:
\begin{equation}
\rho(r'_1)\approx
-\frac{1}{v} \frac{1}{4\pi r^2_{\ast}} \frac{L_{\rm Edd}} {c v_{th}}
\frac{1}{(1-\alpha)}
\left[ 3 (1-\alpha) k\Gamma_D \right]^{1/\alpha}
[{r'_1}^{-1}f'(r'_1) f(r'_1)^\frac{1-\alpha}{\alpha}].
\end{equation}
Finally combining, eqs A2 and A12 we find that the photoionization
parameter is:
\begin{equation}
\xi(r'_1)_{\tau_{\rm X}=0} \approx - (4\pi m_p \mu c v_{th})  x f_{\rm X} (1-\alpha)
\left[ 3 (1-\alpha) k\right]^{-1/\alpha}
\Gamma_D^\frac{\alpha-1}{\alpha}
[{r'_1}f'(r'_1) f(r'_1)^\frac{1-\alpha}{\alpha}]^{-1} v.
\end{equation}
Thus the photoionization parameter, in the optically thin case, 
scales with $x$, $f_{\rm X}$, $\Gamma_D$, $v$, and $r'$ as follows:
\begin{equation}
\xi(r'_1)_{\tau_{\rm X}=0} \propto  x f_{\rm X} \Gamma_D^\frac{\alpha-1}{\alpha}
{r'_1}^{\frac{1}{\alpha}}v. 
\end{equation}

We can estimate the effects of the optical depth on the photoionization
parameter via the X-ray  optical depth along the radial direction:
\begin{equation}
\tau_{\rm X}(r'_1, r'_2)= r_\ast \int_{r'_1}^{r'_2} {\rho(r')} \kappa_{\rm X}
(r') dr',
\end{equation}
where $\kappa_{\rm X}$ is the absorption coefficient representative
for the X-ray range of the photon energy. Evaluation of the 
the X-ray opacity requires detailed photoionization calculations
because $\kappa_{\rm X}$ depends on 
the luminosity and  spectral energy distribution of the ionizing radiation,
the chemical abundances,  and the column density among other things.
For the purpose of our discussion, we consider the local conditions 
at the radius, $r'_1$ (i.e., the radial range 
between $r'_1$ and $r'_1 + \Delta r'$). 
We assume that $\kappa_{\rm X}$ and $\rho$
are constant within $\Delta r'$. 
In other words, we assume that the 
regions at radii smaller than $r'_1$ are optically thin for all radiation.
We  can approximate the characteristic opacity as follows:
\begin{equation}
\tau_{\rm X}(r'_1, r'_1+\Delta r' )\simeq 
\rho(r'_1) \kappa_{\rm X}(r'_1) r_{\ast} \Delta r'.
\end{equation}
Note that our characteristic opacity yields an lower limit for 
the optical depth.
Using eq. A12 we can rewrite eq. A16 as
\begin{equation}
\tau_{\rm X}(r'_1, r'_1+\Delta r')\approx \frac{1}{v} \frac{1}{4\pi r_{\ast}} \frac{L_{\rm Edd}} 
{c v_{th}} \frac{1}{(1-\alpha)}
\left[ 3 (1-\alpha) k\Gamma_D \right]^{1/\alpha}
[{r'_1}^{-1}f'(r'_1) f(r'_1)^\frac{1-\alpha}{\alpha}]
 \kappa_{\rm X}(r'_1)  \Delta r',
\end{equation}
Then the X-ray characteristic opacity scales with the radius $r'_1$ as follows:
\begin{equation}
\tau_{\rm X}(r'_1,r'_1+\Delta r') \propto \frac{1}{v} \frac{L_{Edd}}{r_{\ast}} \Gamma_D^{1/\alpha}
 {r'_1}^{-\frac{1-\alpha}{\alpha}} \kappa_{\rm X}(r'_1)\Delta r'.
\end{equation}

\newpage
\section*{ REFERENCES}
 \everypar=
   {\hangafter=1 \hangindent=.5in}

{
  Abbott, D.C. 1982, ApJ, 259, 28

  Balsara, D.S., \& Krolik J.H. 1993, ApJ, 402, 109

  Blair, W.P., Raymond, J.C., Dupree, A.K., Wu, C.-C., Holm, A.V., \&
  Swank, J. 1984, ApJ, 278, 270

  Begelman, M.C., \& McKee  C.F. 1983, ApJ, 271, 89

  Begelman, M.C., McKee C.F., \& Shields, G.A., 1983, 271, 30 (BMS) 

  Blandford, R.D., Payne, D.G. 1982, MNRAS, 199, 883

  Blondin, J.M. 1994, ApJ, 435, 756

  Blondin, J.M., Kallman, T.R., Fryxell, B.A., \& Taam, R.E. 1990, 
  ApJ, 356, 591

  Castor, J.I., Abbott, D.C.,  Klein, R.I. 1975, ApJ, 195, 157 (CAK)

  Cordova, F.A.,  Mason K.O. 1982, ApJ, 260, 716

  Cunnigham C. 1976 ApJ, 208, 534

  Drew, J.E. 1987, MNRAS, 224, 595

  Drew, J.E., Proga, D. 2000, NewA Rev., 44, 21 

  Dubus, G., Lasota, J.-P., Hameury, J.-M., \& Charles, P. 1999, 
  MNRAS, 303, 139

  Frank, J., King, A.R., \& Raine, D.J. 1985, Accretion Power in Astrophysics,
  (Cambridge: Cambridge University)
 
  Gayley, K.G. 1995, ApJ, 454, 41

  Icke, V. 1981, ApJ, 247, 152

  Kallman, T.R. Boroson B., \& Vrtilek, S.D. 1998, ApJ, 502, 441

  Kallman, T.R., Raymond, J.C. \& Vrtilek, S.D. 1991, ApJ, 370, 717

  Ko, Y.-K., \& Kallman, T.R. 1991, ApJ, 364, 721

  Ko, Y.-K., \& Kallman, T.R. 1994, ApJ, 431, 273

  K\"onigl A., Ruden S.P., 1993, in {\sl Protostars and Planets III},
  eds. E. H. Levy \& J. I. Lunine, Tucson: U. Arizona Press, 641
  
  Mauche, C.W., Raymond, J.C. 2000, ApJ, 541, 924

  McClintock, J.E., London, R., Bond, H., \& Grauer, A. 1982, ApJ, 258, 245 

  Melia, F., Zylstra G.J., \& Fryxell, B. 1991, ApJ, 377, L101 

  Melia, F. \& Zylstra G.J., 1992, ApJ, 398, L53

  Meyer, F. \& Meyer-Hoffmeister, E. 1982, A\&A, 104, L10
 
  Murray, S.D., Castor, J.I., Klein, R.I., \& McKee, C.F. 1994, ApJ, 435, 631

  Murray, N., Chiang, J., Grossman, S.A., \& Voit, G.M. 1995, ApJ, 451, 498 

  Ostriker, E.C, McKee C.F., \& Klein R.I, 1991, ApJ, 377, 593

  Owocki, S.P., Castor J.I., \& Rybicki, G.B. 1988, ApJ, 335, 914

  Pereyra, N.A., Kallman, T.R. \& Blondin, J.M. 2000, ApJ, 532, 563

  Pringle, J.E. 1981, ARAA, 19, 137

  Proga, D. 1999, MNRAS, 304, 938

  Proga, D. 2000, ApJ, 538, 684

  Proga, D., \& Kallman, T.R. 2001, in preparation

  Proga, D., Stone J.M., \& Drew J.E. 1998, MNRAS, 295, 595 (PSD~98)

  Proga, D., Stone J.M., \& Drew J.E. 1999, MNRAS, 310, 476 (PSD~99)

  Proga, D., Stone J.M., \& Kallman, T.R. 2000, ApJ, 543, 686 (PSK)

  Puls, J., Springmann, U., \& Lennon, M. 2000, A\&AS 141, 23

  Raymond, J.C. 1993, ApJ,  412,  267

  Pacharintanakul, P. \& Katz J. 1980, ApJ, 238, 985

  Shakura N.I., \& Sunyaev R.A. 1973 A\&A, 24, 337

  Stevens, I.R.  1991, ApJ, 379, 310

  Stevens, I.R., \& Kallman T.R. 1990, ApJ, 365, 321

  Stone, J.M., Norman, M.L. 1992, ApJS, 80, 753

  van Paradjis J., 1983, in Accretion-Driven Stellar X-ray Sources, eds.,
    W.H.G. Lewin and E.P.J. van de Heuvel 
   (Cambridge University Press: Cambridge), p. 189

  Vitello, P.A.J, \&  Shlosman, I. 1988, 327, 680
 
  Vrtilek, S.D., Raymond, J.C., Garcia, M.R., Verbunt, F.W., \& Hasinger
 G.R. 1990, A\&A, 235, 162

  Vrtilek, S.D., McClintock, J.E., Seward, F.D., Kahn, S. M., 
  \& Wargelin, B. J. 1991a, ApJS, 76, 1127

  Vrtilek, S.D., Penninx, W., Raymond, J.C., Verbunt, F., Hertz, P.,
   Wood, K., Lewin, W.H.G., \& Mitsuda, K. 1991b, 376, 278 

  Vrtilek, S.D., Soker, N., \& Raymond, J.C. 1993, 404, 696

  Vrtilek, S.D., Mihara, T., Primini, F.A., Kahabka, P., Marshall, H., 
  Agerer, F., Charles, P.A., Cheng, F.H., Dennerl, K., La Dous, C., 
  Hu, E.M., Rutten, R., Serlemitsos, P., Soong, Y., Stull, J., Truemper, J., 
  Voges, W., Wagner, R. M., \& Wilson, R. 1994, ApJ, 436, L9

  Woods D.T., Klein, R.I., Castor J.I., McKee, C.F., \& Bell, J.B. 1996, 
  ApJ, 461, 767

  White, N.E., \& Holt, S.S. 1982, ApJ, 257, 318

  Withbroe, G. 1971, in The Menzel Symposium on Solar Physics, Atomic Spectra,
          and Gaseous Nebulae, ed. K.B. Gebbie (NBS Spec. Pub. 353; Washington,
          D.C.: NBS), p. 127

}

\eject

\newpage

\begin{figure}
\begin{picture}(180,530)
\put(140,0){\includegraphics{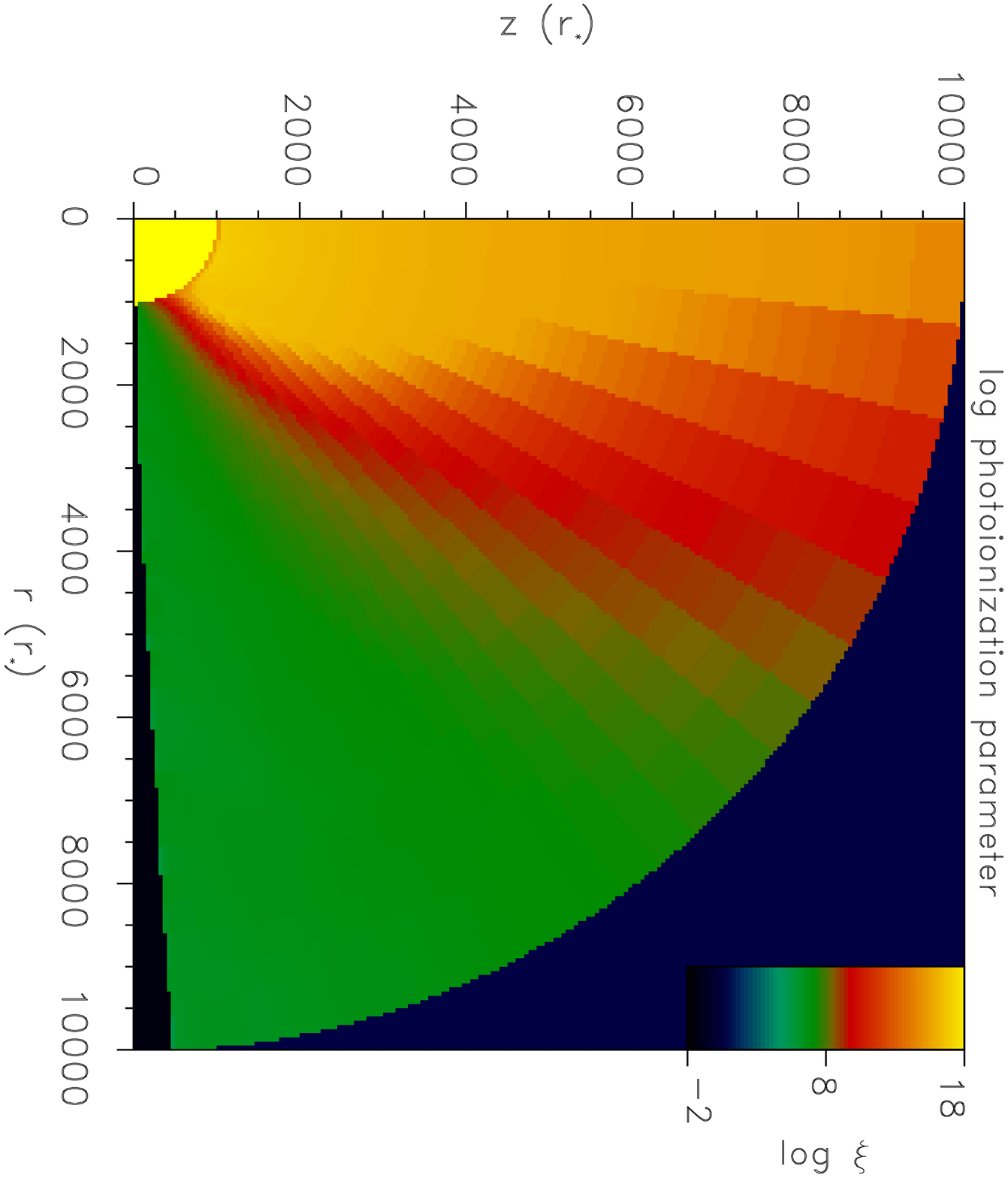}}
\put(140,210){\includegraphics{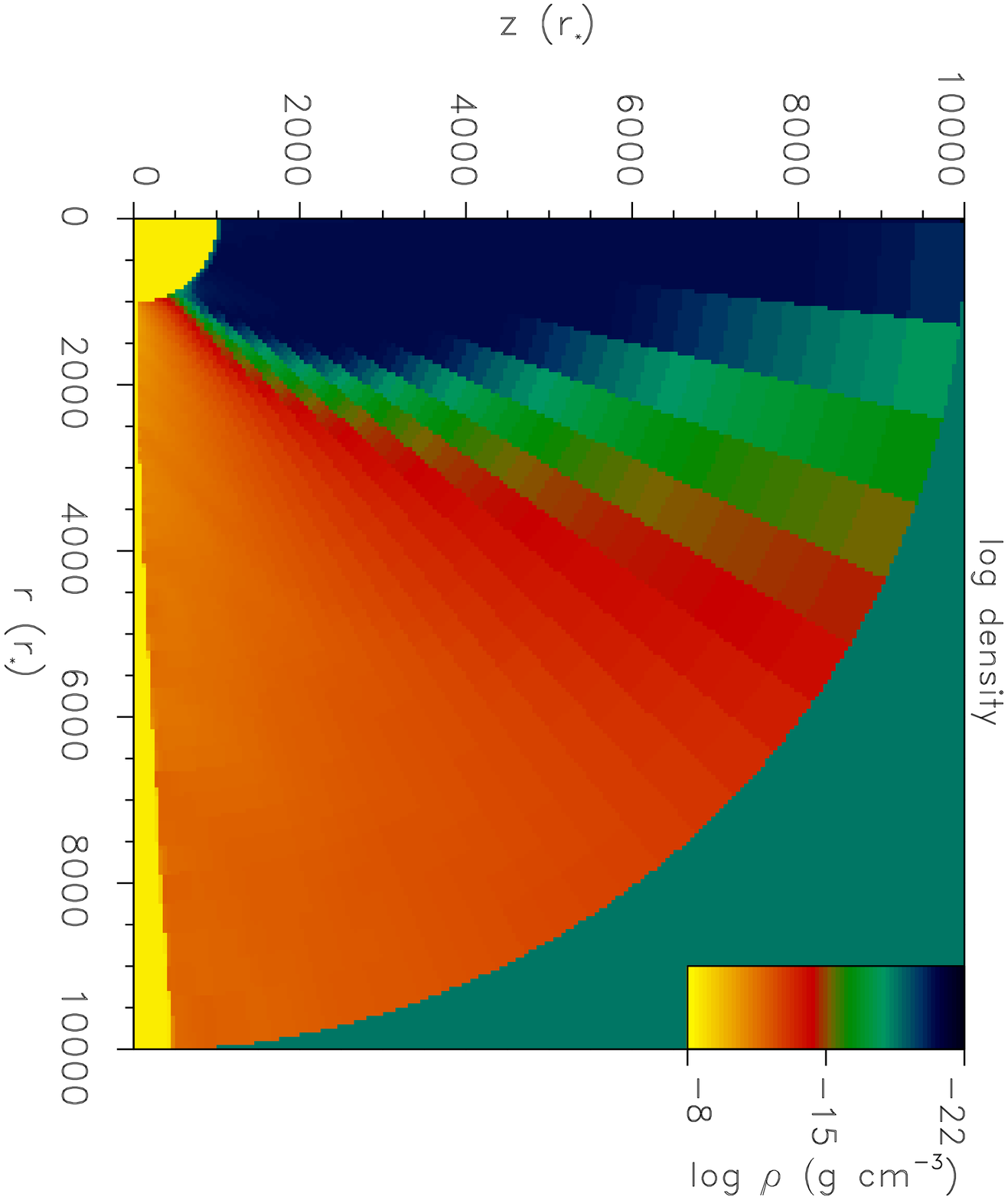}}
\put(230,0){\includegraphics{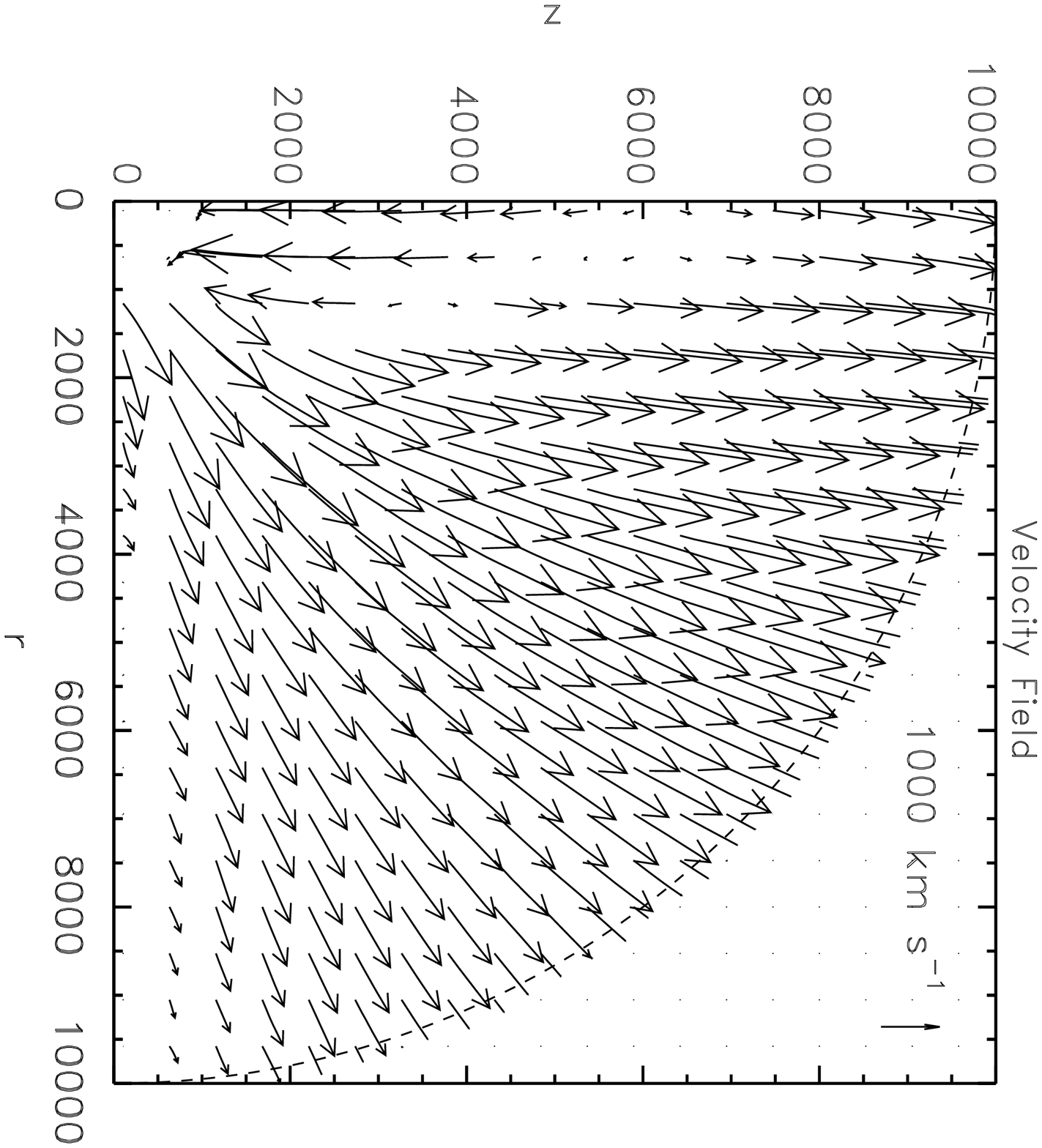}}
\put(230,210){\includegraphics{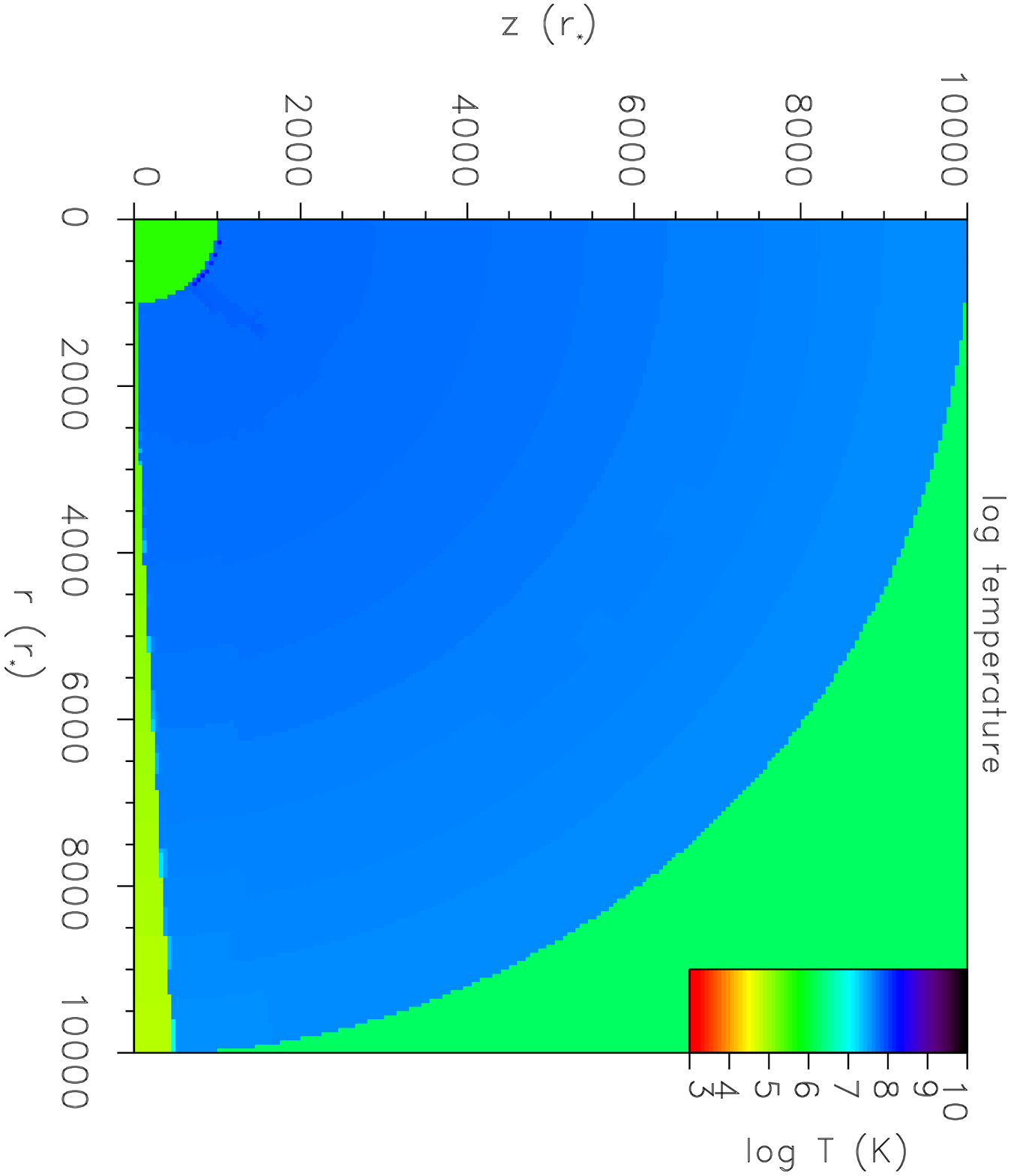}}
\end{picture}
\caption{The top left panel is a color density map of a disk wind
model H2, described in the Section~3. The top right panel is a color
gas temperature map of the model while the bottom left panel
is a color photoionization parameter map. Finally, the bottom right
panel is a map of the velocity field (the poloidal component only). 
In all panels the rotation axis of the disk is along the left hand vertical 
frame, while the midplane of the disk is along the lower horizontal frame.
}
\end{figure}

\begin{figure}
\begin{picture}(600,500)
\put(0,0){\includegraphics{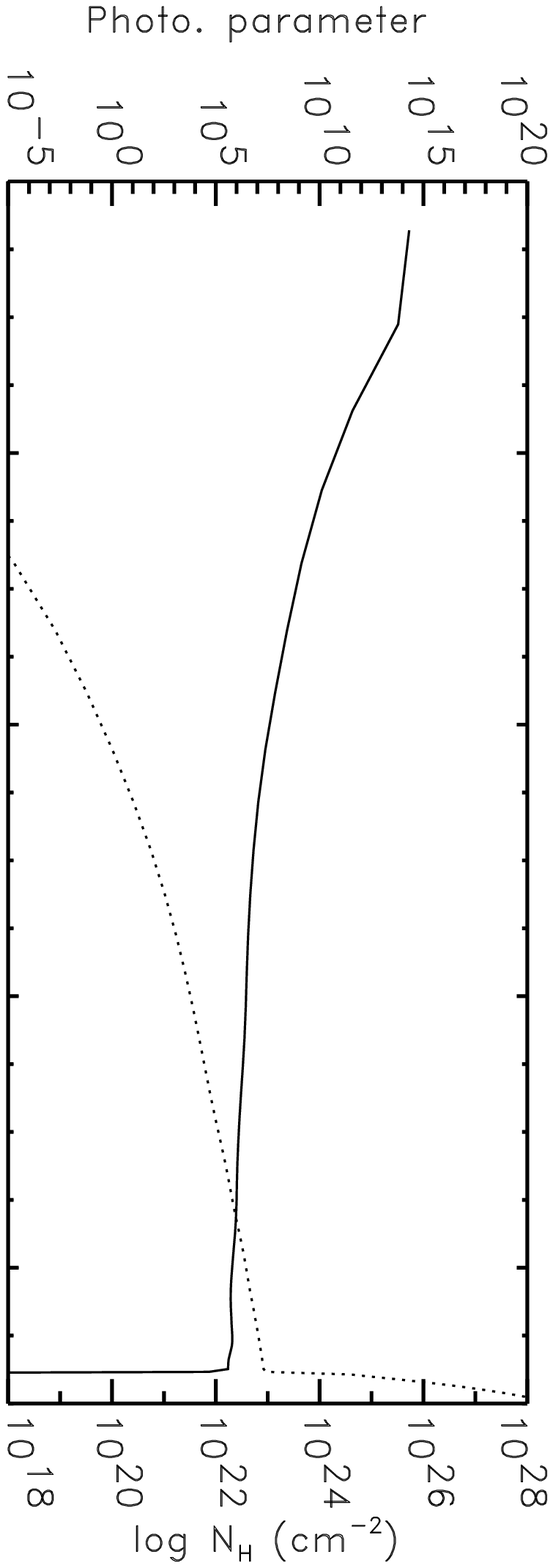}}

\put(0,0){\includegraphics{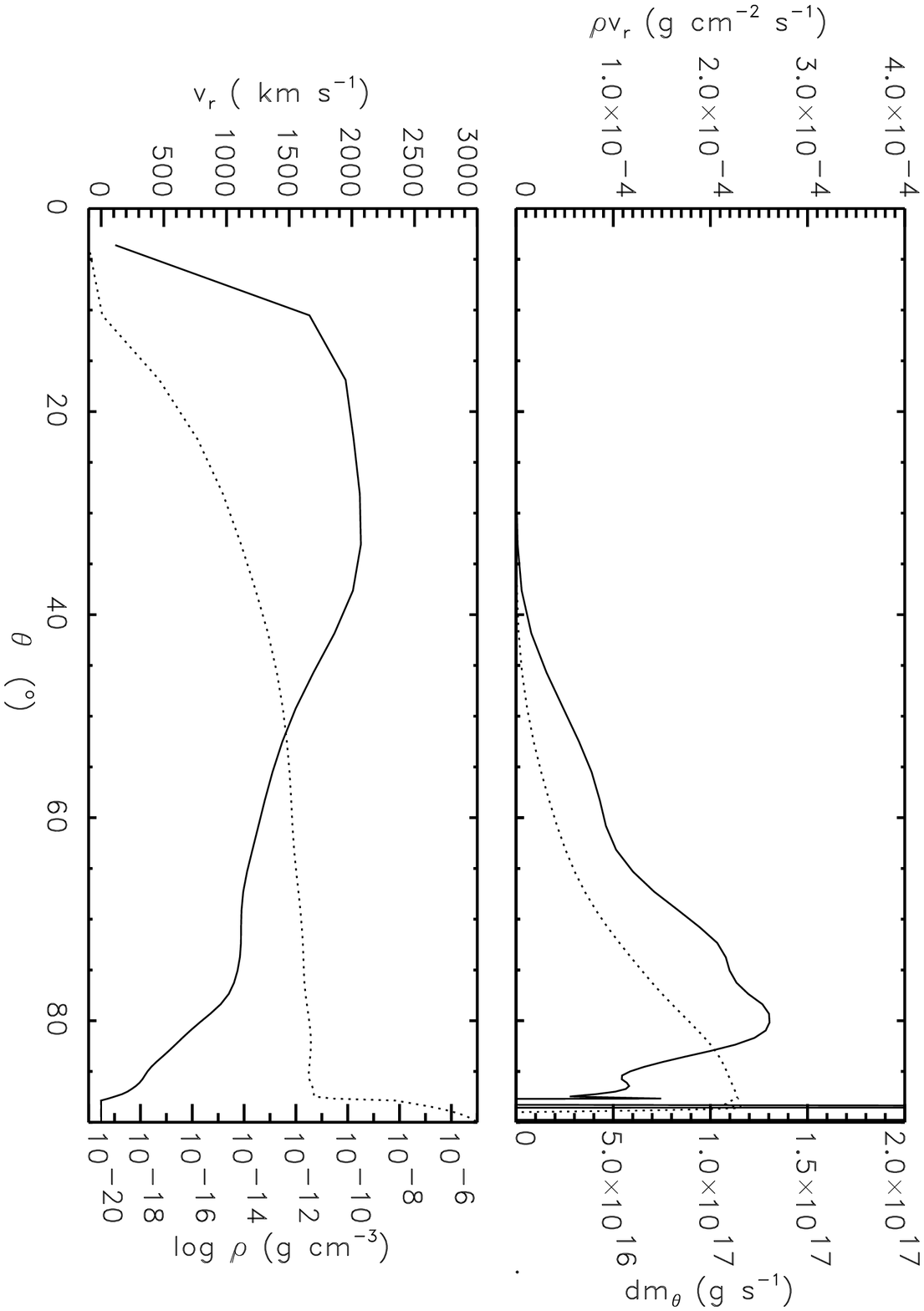}}
\end{picture}
\caption{Quantities at the outer boundary in the model H2.
The ordinate on the left hand side of each panel refers to the solid
line, while the ordinate on the right hand side refers to the dotted
line. 
The column density, $\rm N_H$ is calculated  along the radial direction.}
\end{figure}

\begin{figure}
\begin{picture}(180,530)
\put(140,0){\includegraphics{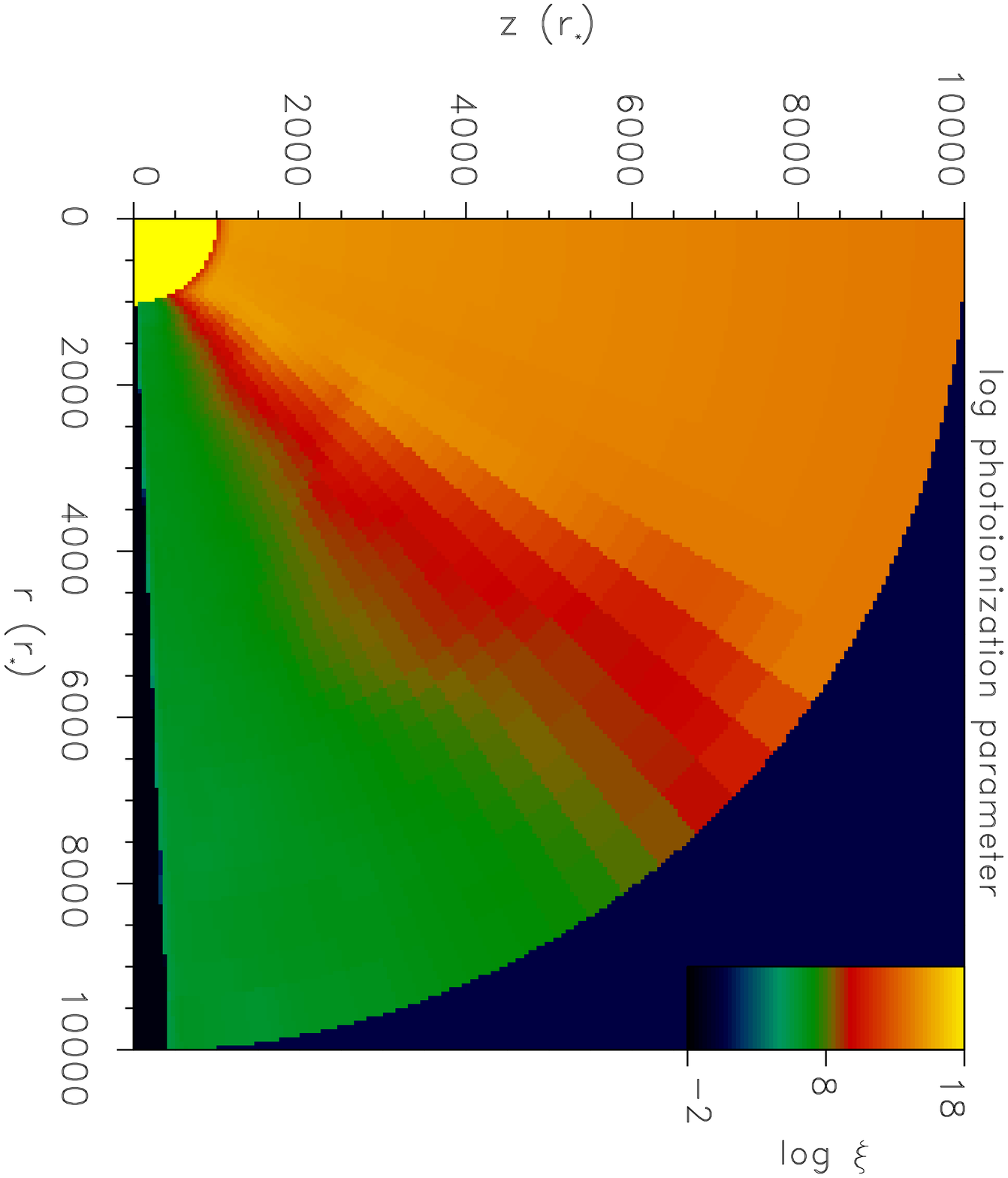}}
\put(140,210){\includegraphics{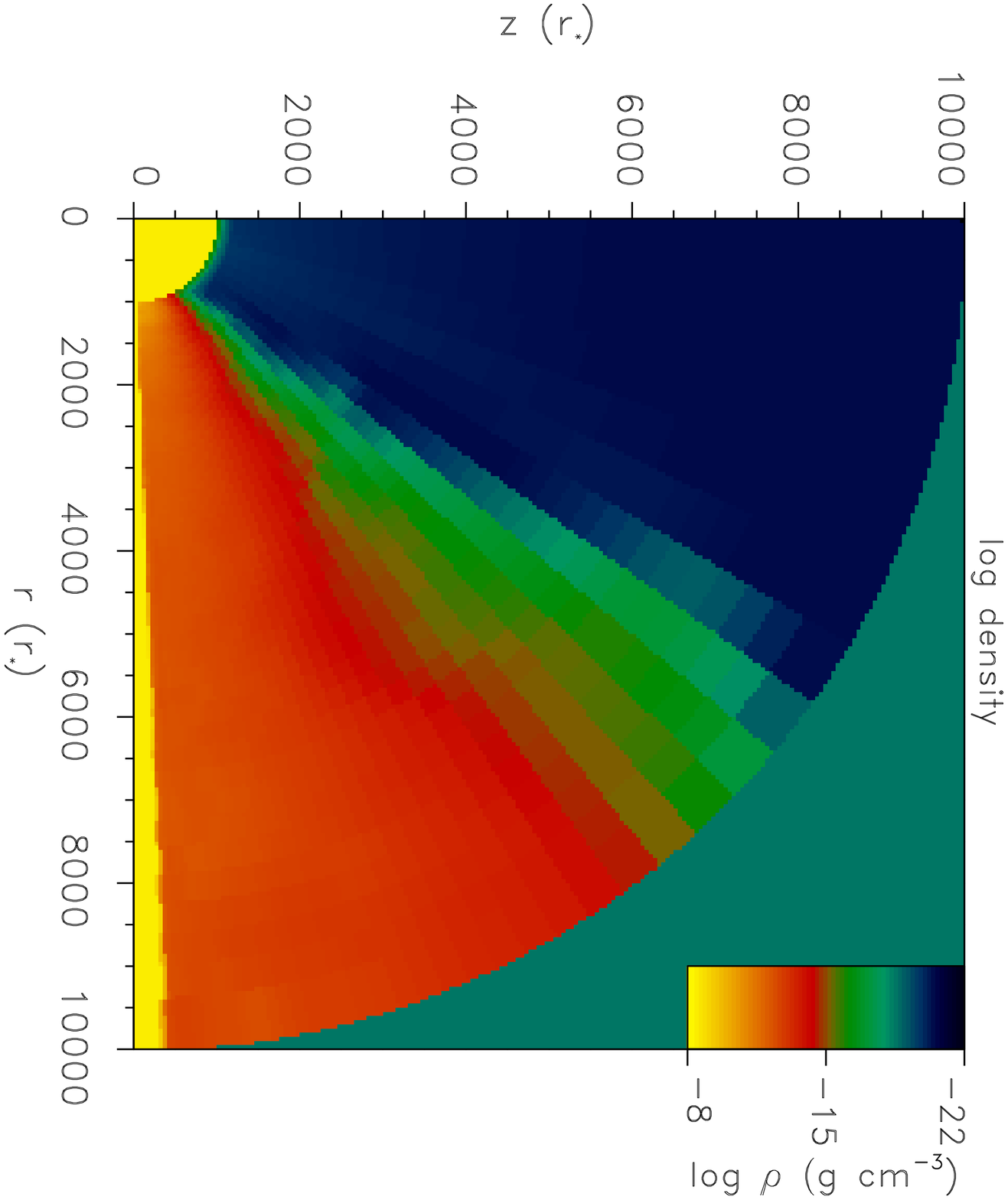}}
\put(230,0){\includegraphics{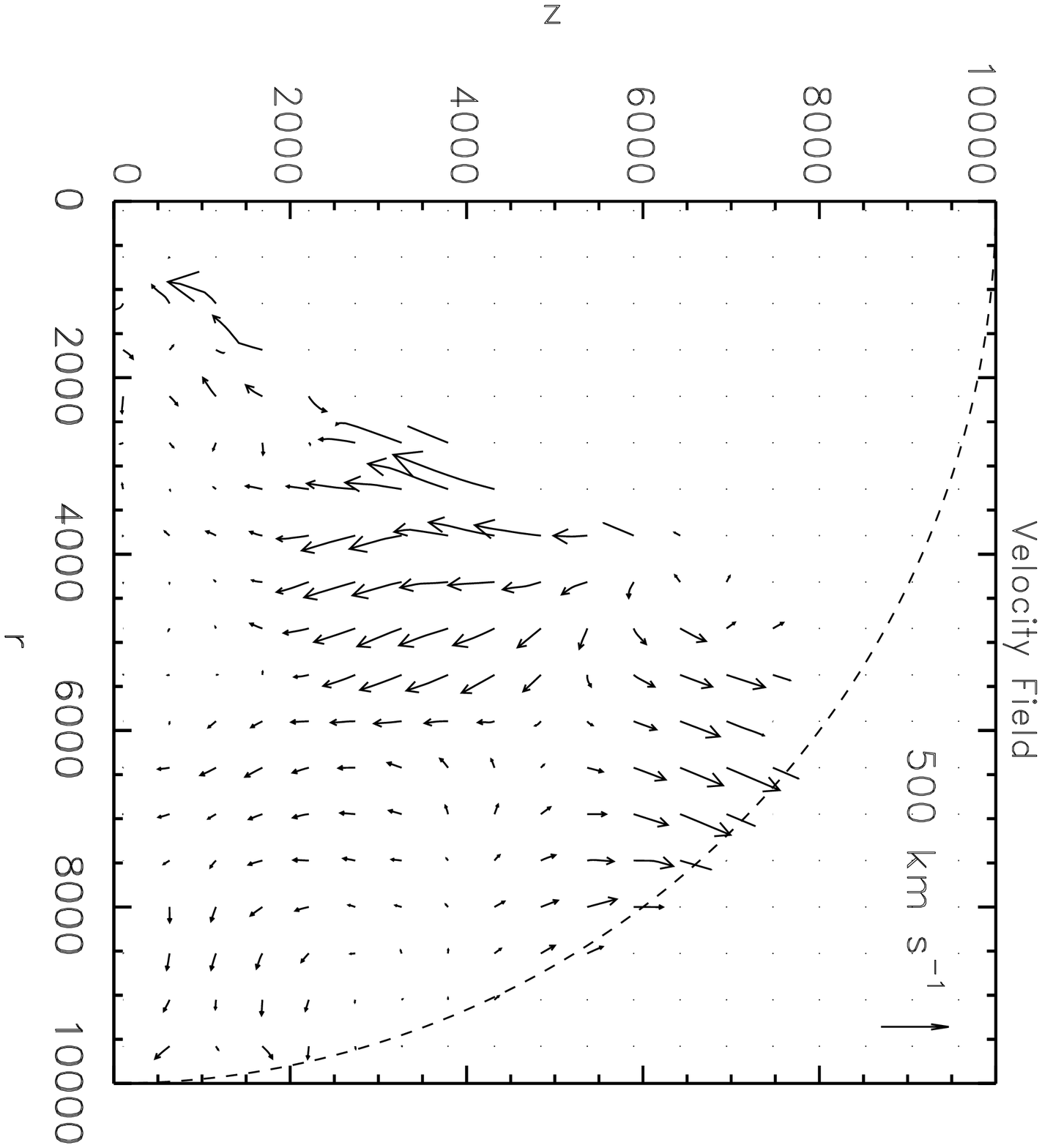}}
\put(230,210){\includegraphics{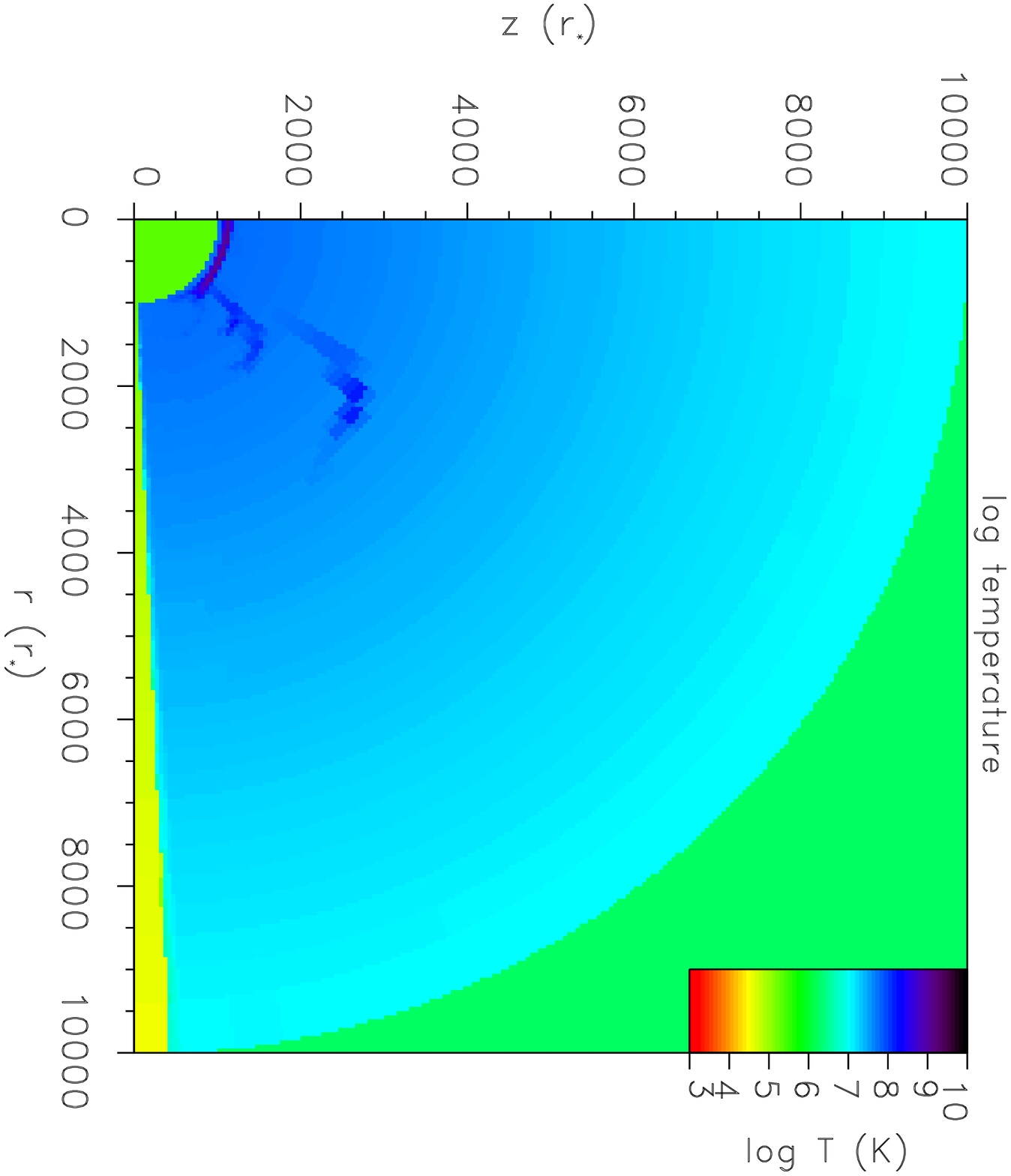}}
\end{picture}
\caption{ As in Figure~1 but for model~L2.Note that we suppress velocity
in regions of very low density (i.e., $\rho < 10^{-18}~{\rm g~cm^{-2}}$).}
\end{figure}

\begin{figure}
\begin{picture}(600,500)
\put(0,0){\includegraphics{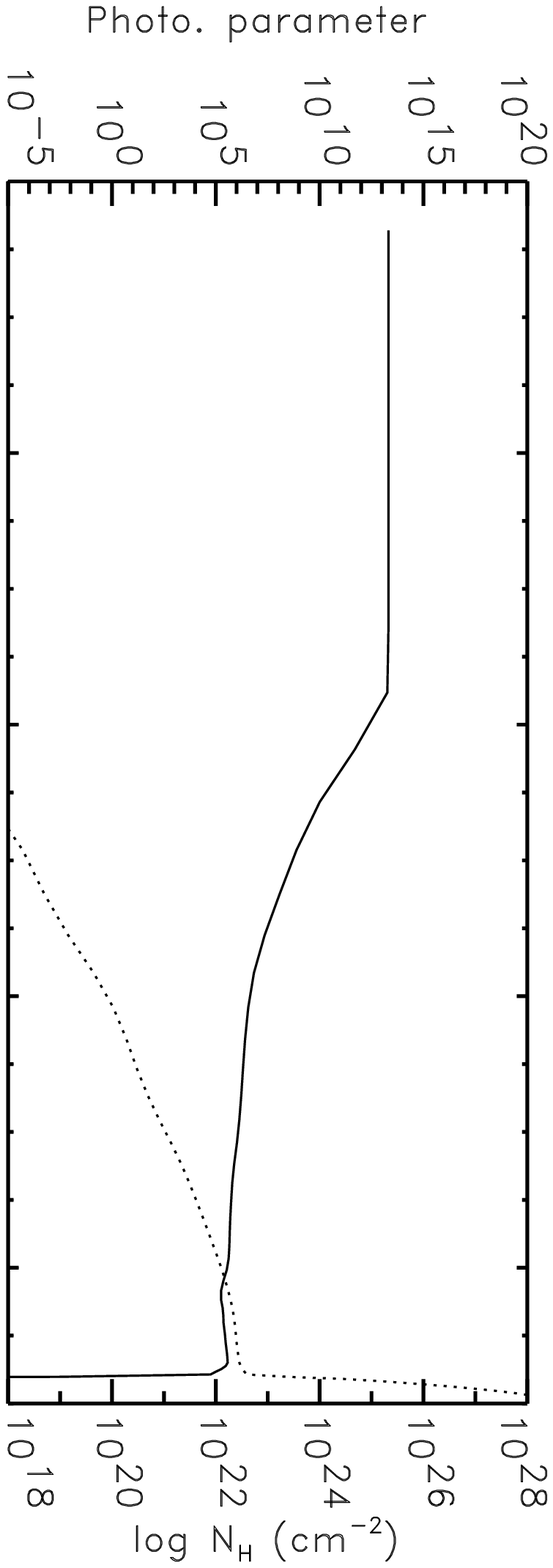}}

\put(0,0){\includegraphics{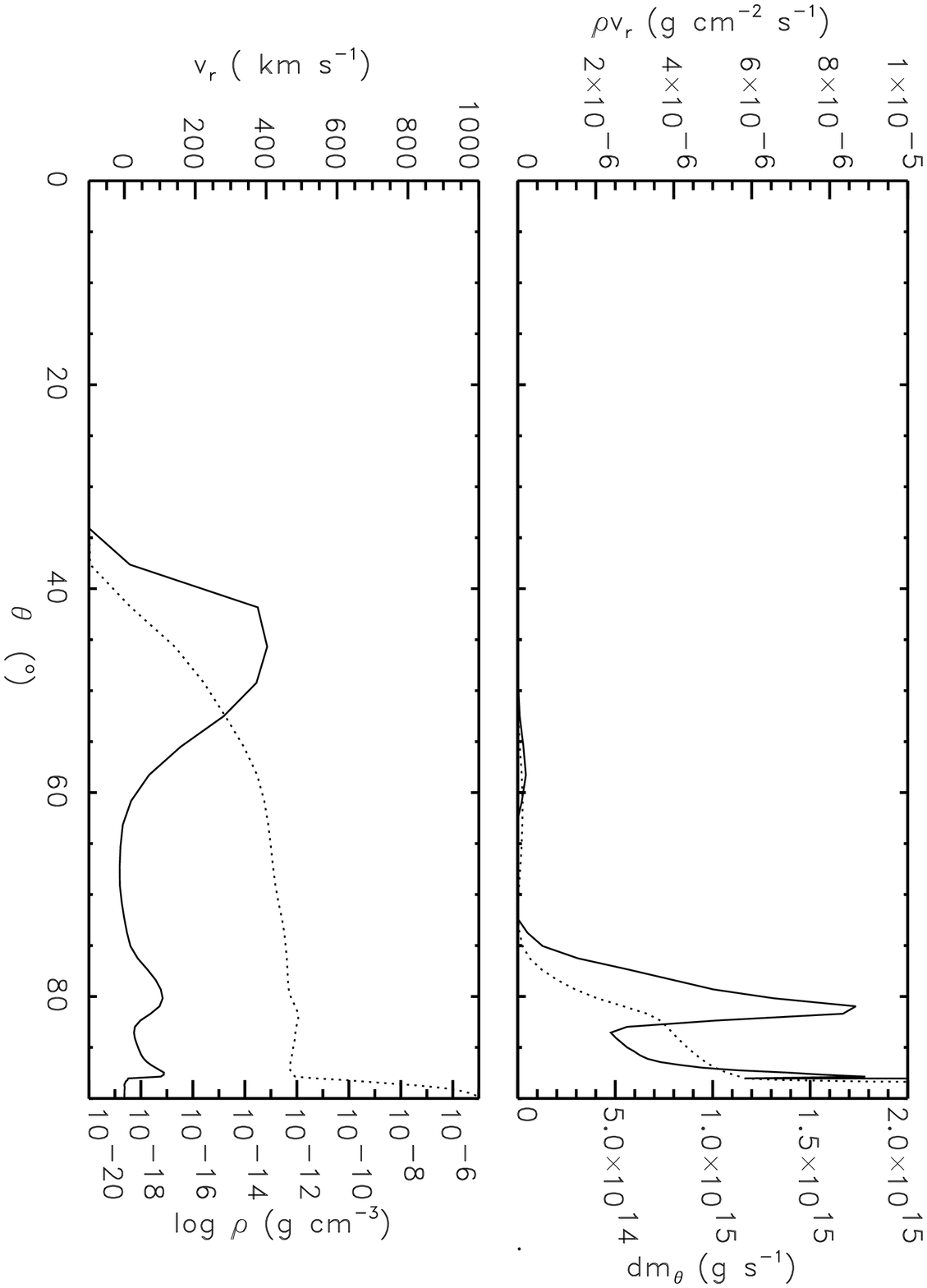}}
\end{picture}
\caption{As in Figure~2 but for model~L2.}
\end{figure}

\begin{figure}
\begin{picture}(600,500)
\put(0,0){\includegraphics{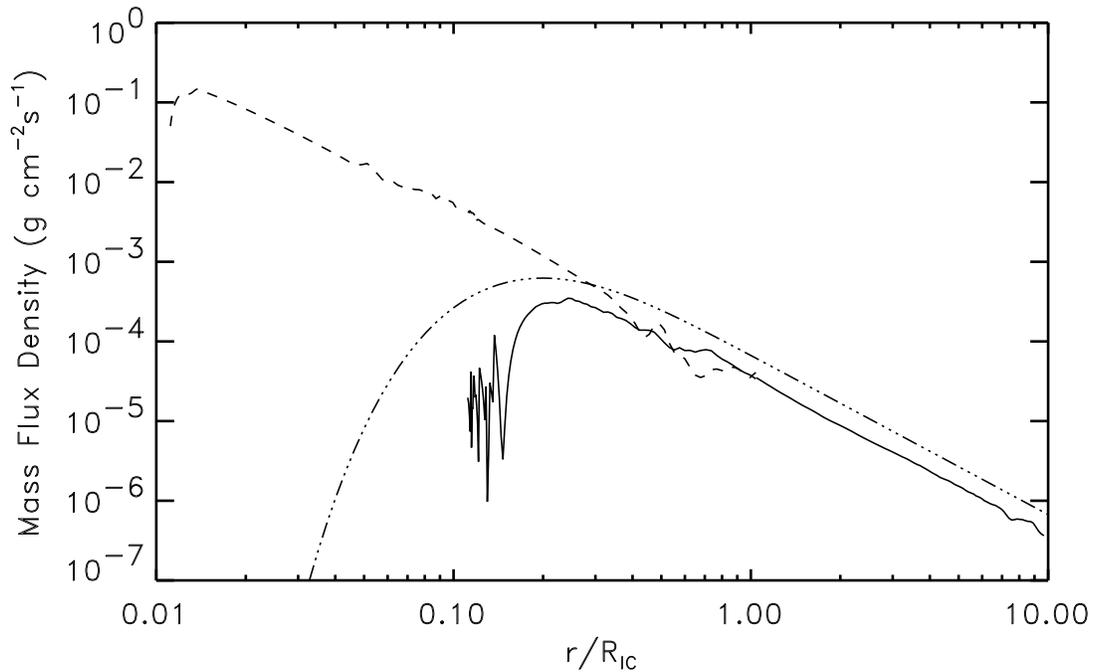}}
\end{picture}
\caption{Mass flux density as a function of radius for the test models.
The solid line represents our 
results for $\Gamma_D=0.667$ with radiation force and X-ray attenuation 
switched off whereas the dashed line represents our results with radiation 
force switched on (in both cases $\kappa_X=\kappa_{UV}=0$).
The mass flux density is measured along a fiducial surface at
a fixed $\theta$ of $87^o$.
For comparison,
the figure  also shows the analytic fit (the triple-dot dashed line) 
by Woods at al. 1995 to their numerical results (their eq. 5.2).
For the constant $C_0$, in Woods et al.'s expression, we adopted 
value of $10^{-4}$.
}
\end{figure}

\begin{figure}
\begin{picture}(600,500)
\put(0,0){\includegraphics{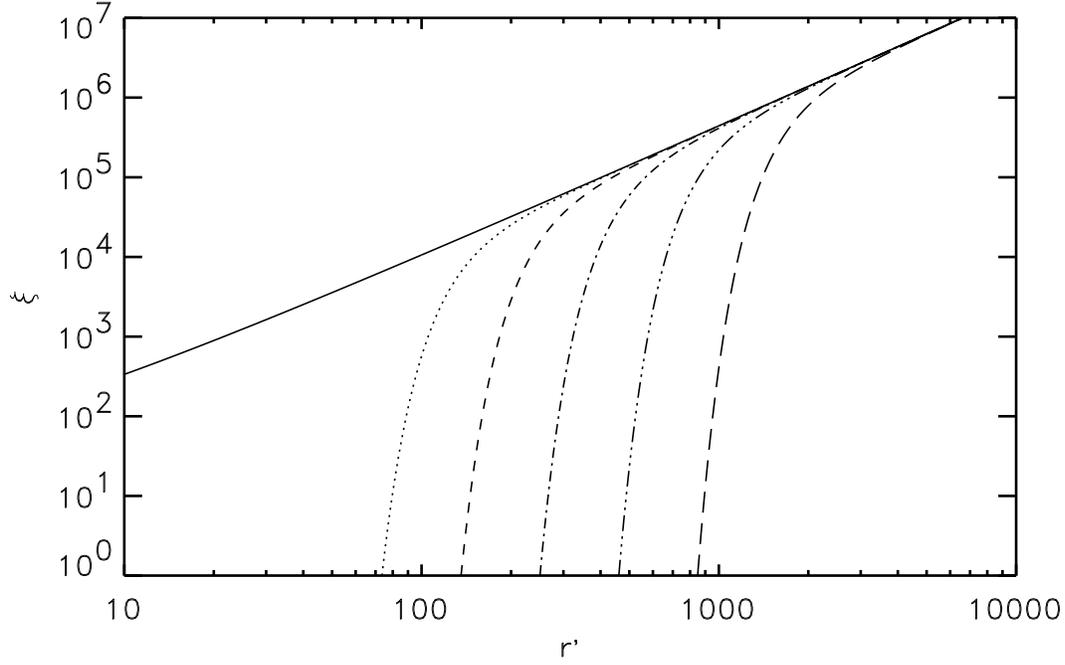}}
\end{picture}
\caption{Estimates of the photoionization parameter as a function
of radius for various X-ray opacities (eq. 15).
The solid line is obtained for $\kappa_{\rm X}=0$, while 
the dotted, dashed, dash-dotted, and long dashed lines
correspond respectively to $\kappa_{\rm X}=4\times10^{-1},
4\times10^{0}, 4\times10^{1}, 4\times10^{2}$ and 
$4\times10^{3}~{\rm g^{-1}~cm ^2}$.}
\end{figure}

\eject
\newpage

\end{document}